\documentclass[acmsmall,screen]{acmart}
\PassOptionsToPackage{table,x11names}{xcolor}
\setcopyright{none}

\usepackage[capitalise,noabbrev]{cleveref}
\usepackage{quantikz}
\usepackage{extarrows}
\usepackage{longtable}
\usepackage{adjustbox}
\usepackage{url}
\usepackage{multicol}
\usepackage{stmaryrd}
\usepackage{amsmath}
\usepackage{physics}
\usepackage{xspace}
\usepackage{todonotes}
\usepackage{minted}
\usepackage{booktabs}
\usepackage{graphicx}
\usepackage{enumerate}
\usepackage{newunicodechar}
\usepackage{xcolor}
\newunicodechar{∥}{\ensuremath{\parallel}}
\newunicodechar{⊕}{\ensuremath{\oplus}}
\newunicodechar{·}{\ensuremath{\cdot}}
\makeatletter
\AtBeginDocument{\def\PYG@tok@err{}}
\AddToHook{env/MintedVerbatim/begin}{\def\PYG@tok@err{}}
\makeatother
\setminted{fontsize=\small,breaklines}

\newtheorem{theorem}{Theorem}
\newtheorem{corollary}[theorem]{Corollary}
\newtheorem{lemma}[theorem]{Lemma}
\theoremstyle{definition}
\newtheorem{definition}[theorem]{Definition}

\newcommand{\Hamilton}{\ensuremath{\textsc{Hamilton}}\xspace}
\DeclareMathOperator*{\ppar}{\parallel}
\newcommand{\sem}[1]{\ensuremath{\left\llbracket #1 \right\rrbracket}}
\newcommand{\usem}[1]{\sem{#1}_U}
\newcommand{\hsem}[1]{\sem{#1}_H}
\newcommand{\snsem}[2]{\sem{#1}_{S_{#2}}}
\newcommand{\ssem}[1]{\snsem{#1}{\infty}}
\newcommand{\pisem}[1]{\sem{#1}_\Pi}
\newcommand{\defeq}{:=}
\newcommand{\id}{0}
\newcommand{\of}{\mathbin{:}}
\newcommand{\sch}{\mathrm{sch}}
\newcommand{\append}{\mathbin{@}}

\begin{document}

\title{A Complete Calculus for Hamiltonian Quantum Programming}

\author{Jacques Carette}
\orcid{}
\affiliation{
  \department{}
  \institution{McMaster University}
  \city{Hamilton}
  \postcode{}
  \country{Canada}
}
\email{carette@mcmaster.ca}

\author{Chris Heunen}
\orcid{0000-0001-7393-2640}
\affiliation{
  \department{School of Informatics}
  \institution{University of Edinburgh}
  \city{Edinburgh}
  \postcode{EH8 9AB}
  \country{United Kingdom}
}
\email{chris.heunen@ed.ac.uk}

\author{Robin Kaarsgaard}
\orcid{0000-0002-7672-799X}
\affiliation{
  \department{Centre for Quantum Mathematics (QM)}
  \department{Centre for Formal Methods and Future Computing (FORM)}
  \institution{University of Southern Denmark}
  \city{Odense}
  \postcode{5230}
  \country{Denmark}
}
\email{kaarsgaard@imada.sdu.dk}

\author{Neil J. Ross}
\orcid{0000-0003-0941-4333}
\affiliation{
  \department{Department of Mathematics and Statistics}
  \institution{Dalhousie University}
  \city{Halifax}
  \postcode{B3H 4R2}
  \country{Canada}
}
\email{neil.jr.ross@dal.ca}

\author{Amr Sabry}
\orcid{0000-0002-1025-7331}
\affiliation{
  \department{Department of Computer Science}
  \institution{Indiana University}
  \city{Bloomington}
  \postcode{47408}
  \country{USA}
}
\email{sabry@iu.edu}


\begin{abstract}
  Given the forces acting on a quantum particle, a \emph{Hamiltonian}
  models its continuous evolution. Quantum computers evolve
  Hamiltonians to simulate physical systems, to solve optimisation
  problems, and to estimate eigenvalues, and even the popular circuit
  abstraction is compiled to Hamiltonians on the hardware. We
  introduce \Hamilton, a process calculus whose parallel operator is
  the sum of interactions acting simultaneously. Its operational
  semantics is the Lie-Trotter product formula, which recovers a sum
  as the limit of ever finer sequential interleavings; compiling a
  program means halving it to a chosen precision and unrolling the
  interleaving into a circuit. The semantics is fully abstract, the
  expressible Hamiltonians are exactly those whose Pauli coefficients
  have the form $m/2^k$, the precision that successive halving
  reaches, and the equational theory is complete and decidable.  A
  full implementation compiles \Hamilton programs to circuits and
  decides their equivalence, and Agda proofs sanity-check the core
  metatheory.
\end{abstract}


\maketitle

\section{Introduction}\label{s:intro}

Quantum programs are often expressed as circuits.  Across quantum
programming languages, from low-level assembly languages like
OpenQASM~\cite{openqasm} to functional languages like
Quipper~\cite{quipper} to equational calculi like
$\Pi_k$~\cite{sqrtpi,freeq}, a program denotes a sequence of
\emph{gates}: elementary unitary operations applied one after another
to an array of qubits.  Sequential composition is the primitive, and
it makes these languages tractable: a circuit is an instruction
stream, which compiler writers and type theorists know how to reason
about.

Yet one of the most common modes of quantum computation, and the
original motivation for quantum computers~\cite{feynman:simulating},
expresses programs by \emph{Hamiltonians}~\cite{lloyd:simulation}, as
in simulation, optimisation~\cite{farhietal:adiabatic,farhietal:qaoa}
and eigenvalue estimation~\cite{kitaev:phase}. A gate jumps from one
state to another, whereas a Hamiltonian defines the continuous
trajectory between them, of which a gate is a snapshot. Formally, a
Hamiltonian is a self-adjoint matrix $H$, its evolution is
$e^{-itH}$, and the snapshot at a time $t$ is a gate. For example, the
state of a qubit is a pair of amplitudes, written
$\alpha\ket{0} + \beta\ket{1}$, and the Hamiltonian
$X = (\begin{smallmatrix} 0 & 1 \\ 1 & 0 \end{smallmatrix})$ drives
$\ket{0}$ along $\cos t\,\ket{0} - i \sin t\,\ket{1}$, a wave that
oscillates between $\ket{0}$ and $\ket{1}$. At $t = \pi/2$ the snapshot
is the \textsc{not} gate, which swaps the two amplitudes, up to a
global phase $-i$; global phases are unobservable, and we ignore
them from now on.

Hamiltonians become interesting when several act at once. Think of a
Hamiltonian as a force driving a quantum particle: when several forces
act on the particle simultaneously, they add. Take $X$ together with
$Z = (\begin{smallmatrix} 1 & 0 \\ 0 & -1 \end{smallmatrix})$, whose
snapshot at $t = \pi/2$ is the phase-flip gate, taking $\alpha\ket{0}
+ \beta\ket{1}$ to $\alpha\ket{0} - \beta\ket{1}$. In
\cref{fig:staircase}, time spent running $X$ is drawn horizontally in
orange and time spent running $Z$ vertically in blue, so running one
and then the other goes around a square on one side or the other, to
the same corner. Whether the two orders reach the same state depends
on how long each runs. Run in full, they give the gates $XZ$ and $ZX$,
which differ only by an overall sign, so the order is
unobservable. Run halfway, the order becomes observable. Write
$\sqrt{X}$ and $\sqrt{Z}$ for the half runs: $\sqrt{X}$ turns each
amplitude into an even blend of the two, and $\sqrt{Z}$ multiplies the
second amplitude by $i$, halfway around the circle from $1$ to
$-1$. From the initial state $\tfrac{1}{\sqrt{2}}(\ket{0} + \ket{1})$,
the two evolutions proceed as:
\[
  \begin{array}{l@{\qquad}l}
    \tfrac{1}{\sqrt{2}}(\ket{0} + \ket{1})
    \xrightarrow{\;\sqrt{X}\;} \tfrac{1}{\sqrt{2}}(\ket{0} + \ket{1})
    \xrightarrow{\;\sqrt{Z}\;} \tfrac{1}{\sqrt{2}}(\ket{0} + i\ket{1})
    \\[4pt]
    \tfrac{1}{\sqrt{2}}(\ket{0} + \ket{1})
    \xrightarrow{\;\sqrt{Z}\;} \tfrac{1}{\sqrt{2}}(\ket{0} + i\ket{1})
    \xrightarrow{\;\sqrt{X}\;} \ket{0}
  \end{array}
\]
In the first line, $\sqrt{X}$ blends equal amplitudes, which leaves
them equal, and $\sqrt{Z}$ then only turns the phase of the second. In
the second, $\sqrt{Z}$ first turns the amplitudes into $1$ and $i$, and
$\sqrt{X}$ blends these so that their contributions to $\ket{1}$
cancel, leaving $\ket{0}$. The same two steps thus give a fair coin in
one order and a certain outcome in the other. Running the two at once,
the diagonal of \cref{fig:staircase}, is a third evolution, which
interleavings of ever smaller slices approach from either side.

\newcommand{\staircase}[1]{%
\begin{tikzpicture}[scale=2.5, baseline=0pt]
  \pgfmathtruncatemacro\steps{2^#1}%
  \foreach \i in {1,...,\steps}{%
    \pgfmathsetmacro\a{(\i-1)/\steps}%
    \pgfmathsetmacro\b{\i/\steps}%
    \fill[gray!18] (\a,\a) -- (\b,\a) -- (\b,\b) -- cycle;
    \fill[gray!18] (\a,\a) -- (\a,\b) -- (\b,\b) -- cycle;
    \draw[orange!85!black, line width=1pt] (\a,\a) -- (\b,\a);
    \draw[blue!80!black, line width=1pt] (\b,\a) -- (\b,\b);
    \draw[blue!80!black, line width=1pt] (\a,\a) -- (\a,\b);
    \draw[orange!85!black, line width=1pt] (\a,\b) -- (\b,\b);
  }
  \draw[black, line width=.8pt, dashed] (0,0) -- (1,1);
  \node[above, font=\footnotesize] at (0.5,1) {$n=#1$};
  \ifnum#1=0
    \node[below, font=\scriptsize, text=orange!85!black] at (0.5,0) {$X$};
    \node[left, font=\scriptsize, text=blue!80!black] at (0,0.5) {$Z$};
  \fi
\end{tikzpicture}%
}

\begin{figure}
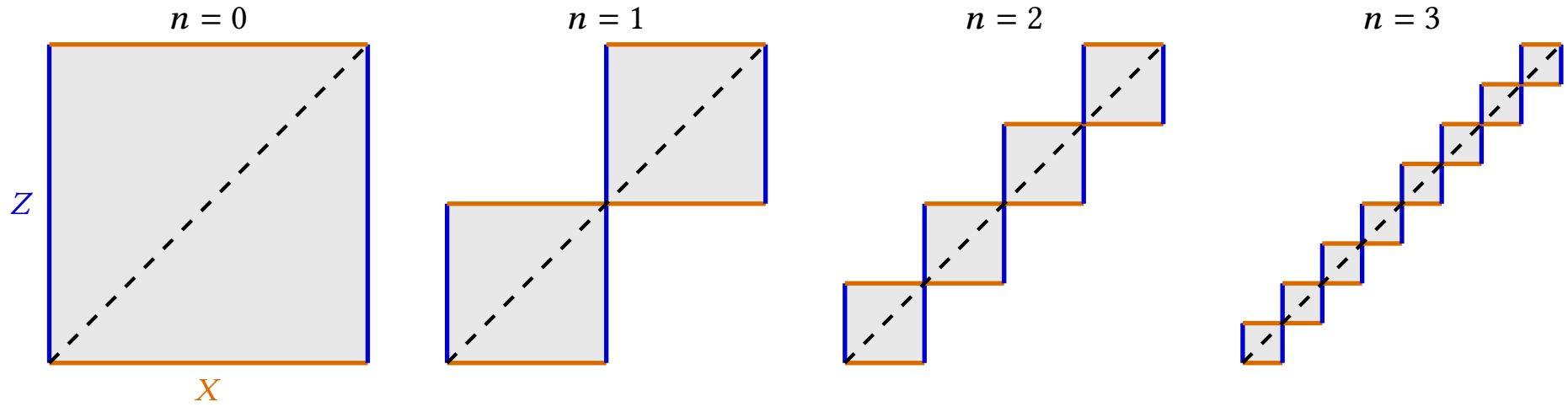

  \centering
  \staircase{0}\hspace{5mm}
  \staircase{1}\hspace{5mm}
  \staircase{2}\hspace{5mm}
  \staircase{3}
  \Description{Four square panels, for levels n equal to 0, 1, 2 and 3.
    Each shows a dashed diagonal from the lower-left to the upper-right
    corner and two staircase paths between the same corners, with steps of
    length one over two to the n; horizontal steps are orange and vertical
    steps blue. One path runs below the diagonal, taking a horizontal step
    first, and the other above it, taking a vertical step first. The
    triangles between each path and the diagonal are shaded; they shrink as
    n grows.}
  \caption{Interleavings of $X$ and $Z$ at levels $n = 0, \ldots, 3$:
    orange steps run $X$, blue steps run $Z$, and the dashed diagonal runs
    both at once. The path below the diagonal runs $X$ first, the path
    above runs $Z$ first. To first order, an interleaving's error is
    proportional to the shaded area between its path and the diagonal,
    which halves at each level.}
  \label{fig:staircase}
\end{figure}

Formally, simultaneous interactions add: their Hamiltonians are summed
in the Lie algebra of the unitary group, while the gates of a circuit
are multiplied in the group itself. Such interactions behave like
concurrent processes. Two processes are independent when the order in
which they run does not affect the outcome; two interactions are
independent in exactly this sense when their Hamiltonians commute, and
otherwise their order shows in the result, as it does for $X$ and~$Z$.
Concurrency theory separates such events from a \emph{schedule} that
orders them~\cite{pratt:geometry,fajstrupetal:directed}, and a single
processor runs concurrent processes by interleaving time slices,
approximating their simultaneous execution better as the slices
shrink. Simultaneous quantum interactions are run the same way. The
Lie-Trotter product formula~\cite{trotter}, the bridge from the
algebra back to the group, writes $e^{A+B}$ as the limit of
$(e^{A/N}e^{B/N})^N$, a round-robin schedule that gives each term a
time slice of $1/N$.

Circuit languages can be patched to reach such sums.
Trotterisation~\cite{lloyd:simulation} writes the round-robin schedule
out as a circuit, and block encodings~\cite{lowchuang:qubitization},
which the language Cobble~\cite{yuan:cobble} compiles, realise a sum
exactly at the cost of ancillary qubits. But, as Milner observed,
concurrency ``requires a new conceptual framework, not just a
refinement of what we find natural for sequential
computing''~\cite{milner:turing}. Rather than patch the circuit model,
we develop that conceptual framework for Hamiltonians, in a language
called \Hamilton. A \Hamilton program composes interactions in
parallel, written $\ppar$, with no order among them, and denotes the
sum of their Hamiltonians. Programs compile to circuits of the
calculus $\Pi_k$~\cite{freeq,sqrtpi}. This is an ideal target, as
$\Pi_k$ extends the calculus $\Pi$ of classical reversible
computing~\cite{jamessabry:pi,10.5555/3089528.3089534} with a square
root of \textsc{not} and with the $2^k$-th roots of unity, phases of
ever finer precision, each the square root of the one before.
Approximation needs square roots, since a Trotter schedule
interleaves ever smaller slices of each interaction, and running an
interaction for half as long takes a square root of its
evolution. \Hamilton adds the matching capability to the process
algebra of independent events: \emph{halving} a process. Compilation
is then the operational semantics: to run a program is to halve it $n$
times, for a chosen level $n$, so that every duration in it is a
\emph{dyadic rational} (a fraction of the form $m/2^k$ for $m \in
\mathbb{Z}$ and $k \in \mathbb{N}$), and to unroll it into a schedule
of $2^n$ Trotter steps, each a $\Pi_k$ circuit.

\paragraph{Outline}
Section~\ref{s:background} gives background on $\Pi_k$ and on
Hamiltonians, through examples. Section~\ref{s:processes} fixes the syntax and
type system of \Hamilton, Section~\ref{s:semantics} its semantics and
universality, and Section~\ref{s:completeness} full abstraction.
Section~\ref{s:implementation} describes the
implementation, and Section~\ref{s:applications} applies it to a
complete example. Section~\ref{s:related} compares \Hamilton with
SimuQ~\cite{pengetal:simuq}, Paulihedral~\cite{li:paulihedral},
Cobble~\cite{yuan:cobble} and equational reasoning about circuits built
from Pauli rotations. Section~\ref{s:conclusion} concludes.

\section{From Hamiltonians to Circuits}\label{s:examples}\label{s:background}

\paragraph{Hamiltonians, Lie algebras, and Lie groups}
The (time-independent) Schrödinger equation describes how physical
systems evolve through time. The specification of the physical system
to evolve is given by a \emph{Hamiltonian} $H$, which is an operator
on the state space of the system satisfying $H^\dagger = H$ (with
$H^\dagger$ the conjugate transpose of $H$). Solutions to this
equation turn out to be given by families of operators $U(t) =
e^{-itH}$ with $t$ the time parameter. These solutions live in a
\emph{Lie group}: a group that is simultaneously a finite-dimensional
real smooth manifold, such that group multiplication and inversion are
smooth. Solutions can be thought of as continuous paths on the surface
of this manifold. When multiplied by $i$, the Hamiltonians themselves
belong to the corresponding \emph{Lie algebra}: a vector space
together with a bilinear map $[-,-]$, the \emph{Lie bracket}, that
satisfies \emph{alternation} $[x,x] = 0$ for all $x$, and the
\emph{Jacobi identity} $[x,[y,z]] + [z,[x,y]] + [y,[z,x]] = 0$ for all
$x,y,z$. Every Lie group is associated with a Lie algebra (and vice
versa) via exponentiation and logarithms. Quintessential examples are
the Lie group $U(2^n)$ of $2^n$-by-$2^n$ unitary matrices, with associated
Lie algebra $\mathfrak{u}(2^n)$.

\paragraph{Free Quantum Computing}
\Hamilton compiles to circuits of $\Pi_k$, the free model of quantum
computing~\cite{freeq,sqrtpi}, so we begin there. Its ancestor $\Pi$
is a calculus of type isomorphisms: in the standard model, types are
finite sets built from $0$, $1$, sums and products, a term is a
permutation of their
elements~\cite{jamessabry:pi,10.5555/3089528.3089534}, and the
calculus is sound and complete for reversible classical
computation. $\Pi_k$ adjoins $V$, a square root of \textsc{not}, and
the scalar phase $\zeta_k = e^{2\pi i/2^k}$, the principal $2^k$-th
root of unity, so that $\zeta_2 = i$. With these extensions, the
language becomes complete for quantum computing with terms denoting
\emph{unitary} matrices. The index $k$ can be any positive integer:
since $\zeta_k = \zeta_{k+1}^2$, each $\Pi_k$ embeds in
$\Pi_{k+1}$. The equational theory of $\Pi_k$ is \emph{fully abstract}
for its unitary semantics~\cite{freeq}: two terms are provably equal
exactly when they denote the same unitary.

\begin{figure}
  \centering
  \begin{tikzpicture}[scale=2]
  \draw[gray!60] (0,0) circle (1);
  \draw[gray!50] (-0.985,-0.059) -- (-0.979,-0.070) -- (-0.972,-0.081) -- (-0.964,-0.091) -- (-0.955,-0.101) -- (-0.945,-0.112) -- (-0.934,-0.122) -- (-0.923,-0.132) -- (-0.910,-0.142) -- (-0.897,-0.151) -- (-0.882,-0.161) -- (-0.867,-0.170) -- (-0.851,-0.180) -- (-0.834,-0.189) -- (-0.816,-0.198) -- (-0.797,-0.206) -- (-0.778,-0.215) -- (-0.758,-0.223) -- (-0.737,-0.231) -- (-0.715,-0.239) -- (-0.692,-0.247) -- (-0.669,-0.254) -- (-0.645,-0.261) -- (-0.621,-0.268) -- (-0.596,-0.275) -- (-0.570,-0.281) -- (-0.544,-0.287) -- (-0.517,-0.293) -- (-0.490,-0.298) -- (-0.462,-0.303) -- (-0.434,-0.308) -- (-0.405,-0.313) -- (-0.376,-0.317) -- (-0.347,-0.321) -- (-0.317,-0.324) -- (-0.287,-0.328) -- (-0.257,-0.331) -- (-0.226,-0.333) -- (-0.195,-0.335) -- (-0.164,-0.337) -- (-0.133,-0.339) -- (-0.102,-0.340) -- (-0.070,-0.341) -- (-0.039,-0.342) -- (-0.007,-0.342) -- (0.025,-0.342) -- (0.056,-0.341) -- (0.088,-0.341) -- (0.119,-0.340) -- (0.150,-0.338) -- (0.181,-0.336) -- (0.212,-0.334) -- (0.243,-0.332) -- (0.274,-0.329) -- (0.304,-0.326) -- (0.334,-0.322) -- (0.363,-0.319) -- (0.393,-0.315) -- (0.421,-0.310) -- (0.450,-0.305) -- (0.478,-0.300) -- (0.505,-0.295) -- (0.532,-0.290) -- (0.559,-0.284) -- (0.585,-0.277) -- (0.610,-0.271) -- (0.635,-0.264) -- (0.659,-0.257) -- (0.682,-0.250) -- (0.705,-0.243) -- (0.727,-0.235) -- (0.748,-0.227) -- (0.769,-0.219) -- (0.789,-0.210) -- (0.808,-0.202) -- (0.826,-0.193) -- (0.843,-0.184) -- (0.860,-0.175) -- (0.876,-0.165) -- (0.890,-0.156) -- (0.904,-0.146) -- (0.917,-0.136) -- (0.929,-0.126) -- (0.941,-0.116) -- (0.951,-0.106) -- (0.960,-0.096) -- (0.968,-0.085) -- (0.976,-0.075) -- (0.982,-0.064) -- (0.988,-0.053) -- (0.992,-0.043) -- (0.996,-0.032) -- (0.998,-0.021) -- (1.000,-0.010);
  \draw[gray!40, densely dotted] (1.000,-0.010) -- (1.000,0.000) -- (0.999,0.011) -- (0.998,0.022) -- (0.995,0.033) -- (0.992,0.043) -- (0.987,0.054) -- (0.982,0.065) -- (0.975,0.075) -- (0.968,0.086) -- (0.960,0.096) -- (0.950,0.106) -- (0.940,0.117) -- (0.929,0.127) -- (0.917,0.137) -- (0.904,0.147) -- (0.890,0.156) -- (0.875,0.166) -- (0.859,0.175) -- (0.842,0.184) -- (0.825,0.193) -- (0.807,0.202) -- (0.788,0.211) -- (0.768,0.219) -- (0.747,0.227) -- (0.726,0.235) -- (0.704,0.243) -- (0.681,0.250) -- (0.657,0.258) -- (0.633,0.265) -- (0.609,0.271) -- (0.583,0.278) -- (0.557,0.284) -- (0.531,0.290) -- (0.504,0.295) -- (0.476,0.301) -- (0.448,0.306) -- (0.420,0.310) -- (0.391,0.315) -- (0.362,0.319) -- (0.332,0.323) -- (0.302,0.326) -- (0.272,0.329) -- (0.241,0.332) -- (0.211,0.334) -- (0.180,0.336) -- (0.149,0.338) -- (0.117,0.340) -- (0.086,0.341) -- (0.054,0.342) -- (0.023,0.342) -- (-0.009,0.342) -- (-0.040,0.342) -- (-0.072,0.341) -- (-0.103,0.340) -- (-0.135,0.339) -- (-0.166,0.337) -- (-0.197,0.335) -- (-0.228,0.333) -- (-0.258,0.330) -- (-0.289,0.327) -- (-0.319,0.324) -- (-0.349,0.321) -- (-0.378,0.317) -- (-0.407,0.312) -- (-0.436,0.308) -- (-0.464,0.303) -- (-0.492,0.298) -- (-0.519,0.292) -- (-0.546,0.287) -- (-0.572,0.281) -- (-0.597,0.274) -- (-0.622,0.268) -- (-0.647,0.261) -- (-0.671,0.254) -- (-0.694,0.246) -- (-0.716,0.239) -- (-0.738,0.231) -- (-0.759,0.223) -- (-0.779,0.215) -- (-0.798,0.206) -- (-0.817,0.197) -- (-0.835,0.188) -- (-0.852,0.179) -- (-0.868,0.170) -- (-0.883,0.161) -- (-0.897,0.151) -- (-0.911,0.141) -- (-0.923,0.131) -- (-0.935,0.121) -- (-0.946,0.111) -- (-0.956,0.101) -- (-0.964,0.090) -- (-0.972,0.080) -- (-0.979,0.069) -- (-0.985,0.059) -- (-0.990,0.048) -- (-0.994,0.037) -- (-0.997,0.027) -- (-0.999,0.016) -- (-1.000,0.005);
  \draw[gray!50] (-1.000,0.005) -- (-1.000,-0.006) -- (-0.999,-0.016) -- (-0.997,-0.027) -- (-0.994,-0.038) -- (-0.990,-0.049) -- (-0.985,-0.059);
  \draw[gray, -latex] (0,0) -- (-0.000,1.300) node[anchor=south, inner sep=1pt, font=\scriptsize] {$Z$};
  \draw[gray, -latex] (0,0) -- (-1.298,-0.078) node[anchor=east, inner sep=1pt, font=\scriptsize] {$X$};
  \draw[gray, -latex] (0,0) -- (0.330,-0.640) node[anchor=north, inner sep=1pt, font=\scriptsize] {$Y$};
  \draw[gray, -latex] (0,0) -- (-0.969,0.866) node[anchor=east, inner sep=1pt, font=\scriptsize] {$X{+}Z$};
  \draw[red, dashed, thick] (-0.547,-0.783) -- (-0.524,-0.790) -- (-0.501,-0.797) -- (-0.477,-0.804) -- (-0.452,-0.810) -- (-0.426,-0.816) -- (-0.400,-0.822) -- (-0.373,-0.827) -- (-0.345,-0.832) -- (-0.317,-0.837) -- (-0.289,-0.841) -- (-0.259,-0.845) -- (-0.230,-0.848) -- (-0.200,-0.851) -- (-0.170,-0.854) -- (-0.139,-0.856) -- (-0.108,-0.857) -- (-0.077,-0.859) -- (-0.046,-0.860) -- (-0.015,-0.860) -- (0.016,-0.860) -- (0.047,-0.859) -- (0.078,-0.859) -- (0.109,-0.857) -- (0.140,-0.856) -- (0.170,-0.853) -- (0.201,-0.851) -- (0.230,-0.848) -- (0.260,-0.845) -- (0.289,-0.841) -- (0.318,-0.837) -- (0.346,-0.832) -- (0.374,-0.827) -- (0.401,-0.822) -- (0.427,-0.816) -- (0.452,-0.810) -- (0.477,-0.804) -- (0.502,-0.797) -- (0.525,-0.790) -- (0.547,-0.782) -- (0.552,-0.766) -- (0.556,-0.749) -- (0.560,-0.730) -- (0.564,-0.711) -- (0.567,-0.690) -- (0.570,-0.667) -- (0.574,-0.644) -- (0.576,-0.620) -- (0.579,-0.594) -- (0.582,-0.568) -- (0.584,-0.541) -- (0.586,-0.512) -- (0.588,-0.483) -- (0.589,-0.453) -- (0.590,-0.422) -- (0.591,-0.391) -- (0.592,-0.359) -- (0.593,-0.326) -- (0.593,-0.292) -- (0.593,-0.258) -- (0.593,-0.224) -- (0.592,-0.189) -- (0.591,-0.154) -- (0.590,-0.119) -- (0.589,-0.083) -- (0.588,-0.047) -- (0.586,-0.011) -- (0.584,0.025) -- (0.582,0.061) -- (0.579,0.097) -- (0.576,0.133) -- (0.574,0.168) -- (0.570,0.204) -- (0.567,0.239) -- (0.564,0.274) -- (0.560,0.308) -- (0.556,0.342) -- (0.552,0.375) -- (0.547,0.408);
  \draw[blue] (-0.547,-0.783) -- (-0.527,-0.789) -- (-0.506,-0.795) -- (-0.485,-0.801) -- (-0.463,-0.807) -- (-0.441,-0.813) -- (-0.418,-0.818) -- (-0.395,-0.823) -- (-0.371,-0.828) -- (-0.346,-0.832) -- (-0.321,-0.836) -- (-0.296,-0.840) -- (-0.292,-0.825) -- (-0.288,-0.809) -- (-0.285,-0.792) -- (-0.281,-0.775) -- (-0.278,-0.756) -- (-0.275,-0.736) -- (-0.272,-0.716) -- (-0.269,-0.694) -- (-0.267,-0.672) -- (-0.264,-0.649) -- (-0.262,-0.625) -- (-0.230,-0.628) -- (-0.197,-0.630) -- (-0.164,-0.633) -- (-0.131,-0.634) -- (-0.098,-0.636) -- (-0.064,-0.637) -- (-0.030,-0.637) -- (0.003,-0.638) -- (0.037,-0.637) -- (0.070,-0.637) -- (0.104,-0.636) -- (0.106,-0.608) -- (0.108,-0.579) -- (0.109,-0.550) -- (0.111,-0.520) -- (0.112,-0.489) -- (0.113,-0.457) -- (0.113,-0.425) -- (0.114,-0.393) -- (0.114,-0.360) -- (0.114,-0.327) -- (0.114,-0.293) -- (0.149,-0.291) -- (0.184,-0.289) -- (0.219,-0.287) -- (0.254,-0.284) -- (0.288,-0.280) -- (0.322,-0.277) -- (0.356,-0.272) -- (0.389,-0.268) -- (0.421,-0.263) -- (0.453,-0.258) -- (0.485,-0.252) -- (0.484,-0.219) -- (0.484,-0.187) -- (0.483,-0.154) -- (0.482,-0.121) -- (0.481,-0.087) -- (0.479,-0.054) -- (0.478,-0.020) -- (0.476,0.014) -- (0.474,0.047) -- (0.471,0.081) -- (0.469,0.115) -- (0.497,0.121) -- (0.524,0.127) -- (0.550,0.133) -- (0.576,0.140) -- (0.601,0.147) -- (0.625,0.155) -- (0.648,0.163) -- (0.671,0.171) -- (0.693,0.179) -- (0.713,0.188) -- (0.733,0.196) -- (0.731,0.223) -- (0.728,0.250) -- (0.725,0.277) -- (0.722,0.303) -- (0.719,0.329) -- (0.716,0.354) -- (0.712,0.380) -- (0.709,0.404) -- (0.705,0.429) -- (0.701,0.452) -- (0.697,0.475);
  \draw[very thick] (-0.547,-0.783) -- (-0.534,-0.782) -- (-0.520,-0.781) -- (-0.506,-0.779) -- (-0.491,-0.776) -- (-0.476,-0.773) -- (-0.461,-0.769) -- (-0.445,-0.765) -- (-0.429,-0.760) -- (-0.413,-0.755) -- (-0.396,-0.749) -- (-0.379,-0.742) -- (-0.362,-0.734) -- (-0.345,-0.727) -- (-0.327,-0.718) -- (-0.309,-0.709) -- (-0.290,-0.700) -- (-0.272,-0.689) -- (-0.253,-0.679) -- (-0.234,-0.667) -- (-0.215,-0.656) -- (-0.195,-0.643) -- (-0.176,-0.631) -- (-0.156,-0.617) -- (-0.137,-0.604) -- (-0.117,-0.589) -- (-0.097,-0.575) -- (-0.077,-0.559) -- (-0.057,-0.544) -- (-0.036,-0.528) -- (-0.016,-0.511) -- (0.004,-0.495) -- (0.024,-0.477) -- (0.044,-0.460) -- (0.064,-0.442) -- (0.084,-0.423) -- (0.104,-0.405) -- (0.124,-0.386) -- (0.144,-0.366) -- (0.164,-0.347) -- (0.183,-0.327) -- (0.203,-0.307) -- (0.222,-0.286) -- (0.241,-0.266) -- (0.260,-0.245) -- (0.279,-0.224) -- (0.297,-0.203) -- (0.316,-0.181) -- (0.334,-0.160) -- (0.351,-0.138) -- (0.369,-0.117) -- (0.386,-0.095) -- (0.403,-0.073) -- (0.419,-0.051) -- (0.436,-0.029) -- (0.451,-0.007) -- (0.467,0.015) -- (0.482,0.037) -- (0.497,0.059) -- (0.511,0.081) -- (0.525,0.103) -- (0.539,0.124) -- (0.552,0.146) -- (0.564,0.168) -- (0.577,0.189) -- (0.589,0.210) -- (0.600,0.231) -- (0.611,0.252) -- (0.621,0.273) -- (0.631,0.294) -- (0.640,0.314) -- (0.649,0.334) -- (0.657,0.354) -- (0.665,0.373) -- (0.673,0.392) -- (0.679,0.411) -- (0.686,0.430) -- (0.691,0.448) -- (0.696,0.466) -- (0.701,0.483);
  \fill (-0.547,-0.783) circle (0.03) node[anchor=north, font=\scriptsize] {start};
  \fill (0.701,0.483) circle (0.03) node[anchor=west, font=\scriptsize] {end};
\end{tikzpicture}
  \Description{A sphere with its equator drawn in grey and axes labelled
    Z pointing up, X pointing left, Y pointing towards the reader, and X
    plus Z between X and Z. A thick arc runs diagonally across the front
    of the sphere, from a point marked start below the equator at the
    lower left to a point marked end above the equator on the right. A
    blue zigzag of eight short turns stays close to the arc and ends near
    the same point. A red dashed path of two long turns swings out to the
    lower right and ends far from it.}
  \caption{The example of \cref{s:background} on the sphere of one-qubit
    states, drawn with $Z$ up, $X$ to the left and $Y$ towards the
    reader. From the marked start, $e^{-i\theta(X+Z)}$ for
    $\theta \in [0, \pi/4]$ turns about the axis of $X+Z$ along a great
    circle (thick), crossing the equator. One turn about $Z$ followed by
    one about $X$ ends far from it (dashed); four alternating turns of a
    quarter the size stay close all the way (blue).}
  \label{fig:bloch}
\end{figure}
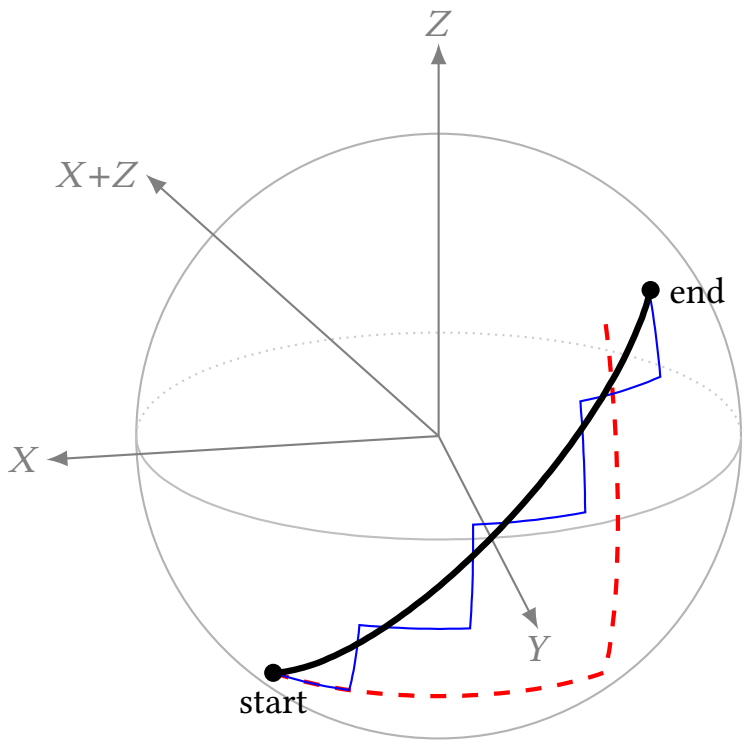

\paragraph{Hamiltonians on the Sphere}
On one qubit, the Hamiltonians of \cref{s:intro} have a geometric
picture. The states of a qubit form a sphere, the \emph{Bloch sphere},
with $\ket{0}$ and $\ket{1}$ at the poles and the states with even
split probabilities $\tfrac{1}{\sqrt{2}}(\ket{0} + e^{i\phi}\ket{1})$
around the equator.  Running the Hamiltonian $Z$ rotates around the
$Z$ axis, i.e., turns the sphere about the axis through the poles,
which changes only the phase $\phi$; running the Hamiltonian $X$ turns
it about an axis through the equator, which moves states between the
poles. Running both at once, $X + Z$ turns the sphere about the axis
halfway between the two, and a state at right angles to that axis
moves along a \emph{great circle}, a circle centred at the centre of
the sphere. One long $Z$-rotation followed by one long $X$-rotation
produces a trajectory that is quite far from the great
circle. Alternating four short turns about $X$ and $Z$ follows that
great circle more closely (\cref{fig:bloch}). In the limit, the
alternation defined by the Lie-Trotter product formula~\cite{trotter}
\begin{equation}\label{eq:trotter}
  e^{-i(A+B)} \;=\; \lim_{n\to\infty}\bigl(e^{-iA/n}\,e^{-iB/n}\bigr)^{n},
\end{equation}
converges to the great circle perfectly.

$X + Z$ is a Hamiltonian, and \cref{fig:bloch} runs it for a chosen
angle, following $e^{-i\theta(X+Z)}$ up to $\theta = \pi/4$. A \Hamilton
program builds the angle into its Hamiltonian instead: a program with
Hamiltonian $H$ runs it for one unit, as the unitary $e^{-2\pi i H}$.
The motion of \cref{fig:bloch}, $e^{-i(\pi/4)(X+Z)}$, is therefore the
program with Hamiltonian $(X+Z)/8$, written with the phases and gates
of $\Pi_k$ as the parallel composition $x \ppar z$ of the two actions
\[
  x = \mathsf{H}(\delta_3 \oplus -\delta_3)\mathsf{H}^\dagger
  \qquad\text{and}\qquad
  z = \delta_3 \oplus -\delta_3.
\]
Here $\delta_3$ is the $1$-by-$1$ Hamiltonian $1/8$, whose unitary is
the $\Pi_3$ phase $\zeta_3$, so $z$, a direct sum, is the
diagonal $Z/8$; conjugating it by the Hadamard gate $\mathsf{H}$ gives
$x$, which is $X/8$; and $\ppar$ adds the two.
The index $3$ is the precision: running an action for half as long
replaces $\delta_3$ by $\delta_4$, a phase of $\Pi_4$.

Every Hamiltonian in \Hamilton has this form. On $n$ qubits it is a
real weighted sum of \emph{Pauli strings}, tensor products of $n$
matrices drawn from $I$, $X$, $Y = iXZ$ and $Z$, one per qubit, and
its program is the parallel composition of its weighted strings. The
weights are dyadic rationals, which are dense in the reals, so an
arbitrary Hamiltonian is matched as closely as desired by refining $k$, and
\cref{thm:universality} shows \Hamilton expresses exactly all Pauli strings
with dyadic weights.

A program runs through its \emph{schedules}, the lists of its actions
in some order. The schedules of $x \ppar z$ are $[x, z]$ and $[z, x]$
(which are lists, not to be confused with the Lie bracket). At level
$n$, a schedule runs each action for $1/2^n$ of the time, so that
$\delta_3$ becomes $\delta_{3+n}$, and repeats $2^n$ times;
\cref{fig:bloch} draws $[z, x]$ at levels $0$ (dashed) and $2$
(blue). The two schedules compile to different circuits at every level
but converge to the same motion, so the theory should equate them. For this, two
rules suffice. The first equates any two orders of a schedule, since
order does not matter in the limit. The second replaces an action by
actions with the same effect at every level, such as $z$ by two copies
of its half, $\delta_4 \oplus -\delta_4$; it asks for equations
between $\Pi_k$ circuits, and the full abstraction of $\Pi_k$ supplies
exactly the true ones. \Cref{s:processes} defines processes, schedules
and these two rules, \cref{s:semantics} gives the semantics and shows
the rules sound, and \cref{s:completeness} shows them complete.

\section{Processes}\label{s:processes}
\newcommand{\exAIdle}{1.000}
\newcommand{\exATurn}{0.026}
\newcommand{\exATurnZero}{1.000}
\newcommand{\exATurnTwo}{0.021}
\newcommand{\exCCnotComposed}{8}
\newcommand{\exCCnotNormal}{4}


This section gives the syntax and types of \Hamilton processes, the
shorthands used in examples, and the rules that equate processes.

\subsection{Syntax and types}\label{s:syntax}

The syntax of \Hamilton follows the structure of a Hamiltonian. Every
Hamiltonian is diagonal in a suitable basis, where it simply assigns a
real number, its rate, to each basis vector; running the Hamiltonian for
one unit of time rotates the phase of each basis vector by its rate,
counted in full circles. Programs are built the same way, from a
diagonal of rates and a basis in which to run it, and several of these
can then run together.

The rates on a diagonal are written as \emph{scalars}, which are chosen
with compilation in mind. The scalar $\delta_k$ runs at rate $1/2^k$, its
opposite $-\delta_k$ at rate $-1/2^k$, and $\id$ stands still.
Compilation cuts a program into ever smaller slices, and a slice half as
long is the same program at half the rate. Since halving $\delta_k$
gives $\delta_{k+1}$, every slice is again made of scalars. Every scalar
also compiles exactly: running $\delta_k$ for one unit of time
multiplies by a $2^k$-th root of unity, which $\Pi_k$ provides. Placing
scalars side by side along a diagonal, with the direct sum $\oplus$,
gives a \emph{bare action}. For instance, the action $z$ of
\cref{s:background} places $\delta_3$ beside $-\delta_3$, rotating the
phase of $\ket{0}$ by an eighth of a circle in one direction and that of
$\ket{1}$ by an eighth in the other; its Hamiltonian is $Z/8$.

A bare action only rotates the phases of the basis vectors. To drive
$\ket{0}$ towards $\ket{1}$, as $X$ does, an \emph{action} runs a bare
action in another basis. It does so by conjugation with a
\emph{transformation}, which switches to the new basis, lets the bare
action run there, and switches back. The action $x$ of
\cref{s:background} runs $z$ in the basis of the Hadamard gate, which
makes its Hamiltonian $X/8$. The transformations are the Clifford
circuits, built from the Hadamard gate, the phase gate, the
controlled-not and the swap by sequential composition and tensor
product, and they are chosen for their fit with Pauli strings. On the
one hand, every Pauli string becomes diagonal, made of $I$ and $Z$
alone, in some Clifford basis, so actions express every Pauli
string. On the other hand, Clifford circuits map Pauli strings to Pauli
strings, so the Pauli coefficients of every process remain dyadic
(\cref{thm:characterisation}) and normal forms decide equivalence
(\cref{s:implementation}). Two changes of basis in a row combine into
one, so an action changes basis only once, around a whole diagonal; and
direct sums combine only bare actions, because changing the basis of
one block of a direct sum and not the other is a controlled operation,
in general outside the Clifford circuits. Every action is therefore a
diagonal of scalars in a single Clifford basis, just like the
diagonalised Hamiltonian we started from.

Finally, several actions run simultaneously in a \emph{process}, whose
Hamiltonian is the sum of theirs; the program $x \ppar z$ of
\cref{s:background} is such a process, with Hamiltonian $(X+Z)/8$.

\begin{figure}
\begin{align*}
  u & \defeq \mathsf{I} \mid \mathsf{H} \mid \mathsf{S} \mid
  \mathsf{CX} \mid \mathsf{SWAP} \mid u \cdot u \mid u \otimes u &
  \text{(transformations)} \\
  d & \defeq \id \mid \delta_k \mid -\delta_k \mid d \oplus d &
  \text{(bare actions)} \\
  a & \defeq d \mid uau^\dagger & \text{(actions)} \\
  p & \defeq a \mid p \ppar p & \text{(processes)}
\end{align*}
\begin{gather*}
  \frac{}{\mathsf{I} \of 2} \qquad \frac{}{\mathsf{H} \of 2} \qquad
  \frac{}{\mathsf{S} \of 2} \qquad \frac{}{\mathsf{CX} \of 4} \qquad
  \frac{}{\mathsf{SWAP} \of 4} \\[0.5\baselineskip]
  \frac{u_1 : n \quad u_2 : m}{u_1 \otimes u_2 : nm} \qquad
  \frac{u_1 : n \quad u_2 : n}{u_1 \cdot u_2 : n} \\
  \frac{}{\id \of 1} \qquad \frac{}{\delta_k \of 1} \qquad
  \frac{}{-\delta_k \of 1} \\
  \frac{d \of n \quad d' \of m}{d \oplus d' \of n+m} \qquad
  \frac{a \of n \quad u \of n}{uau^\dagger \of n} \qquad
  \frac{p \of n \quad p' \of n}{p \ppar p' \of n}
\end{gather*}
\Description{A grammar and typing rules. Transformations u are I, H, S, CX,
  SWAP, sequential composition u dot u and tensor product u tensor u. Bare
  actions d are id, delta k, minus delta k and direct sums d plus d. Actions
  are bare actions or conjugations u a u dagger, and processes are actions or
  parallel compositions p parallel p. The typing rules give I, H and S type 2,
  CX and SWAP type 4 and the scalars type 1; tensor products multiply types,
  direct sums add them, and sequential composition, conjugation and parallel
  composition require equal types.}
\caption{The syntax and type system of $\Hamilton$. Types are positive
  integers $n,m \in \mathbb{Z}_+$ corresponding to the dimension of the
  space acted on by the process. Among the transformations, $\mathsf{H}$
  is the Hadamard gate and $\mathsf{S}$ the phase gate
  $\mathrm{diag}(1, i)$.}
\label{fig:syntax}
\end{figure}
\Cref{fig:syntax} gives the formal grammar, together with a type system
that keeps dimensions consistent. The type of a term is the dimension
of the space it acts on: a scalar has type one, a direct sum adds the
types of its parts, and conjugation and parallel composition require
their parts to have equal types. Transformations act on qubits, so
their types are powers of two, and we focus on processes of these
types, which act on registers of qubits.

Halving extends from scalars to every process, running it for $1/2^n$
of its duration.

\begin{definition}
  The \emph{$n$-way halving} $p/2^n$ of a process $p$ is given by
  $\id/2^n = \id$, $\delta_k/2^n = \delta_{k+n}$ and
  $-\delta_k/2^n = -\delta_{k+n}$, and commutes with the other
  constructs: $(d \oplus d')/2^n = d/2^n \oplus d'/2^n$,
  $(uau^\dagger)/2^n = u(a/2^n)u^\dagger$ and
  $(p \ppar p')/2^n = p/2^n \ppar p'/2^n$.
\end{definition}

\subsection{Derived forms}\label{s:derived}
The core syntax is minimal, and examples use a few shorthands that
expand into it. We write $\id_n$ for the action $\id \oplus \cdots
\oplus \id$ of type $n$, and $\ppar_{i=1}^n p_i$ for
$p_1 \ppar \cdots \ppar p_n$, which needs no parentheses since $\ppar$
is associative (\cref{prop:abelian}). Conjugating a composite process
conjugates each part, $u(\ppar_i p_i)u^\dagger = \ppar_i u p_i
u^\dagger$. Repetition $n \star p = \ppar_{i=1}^n p$, combined with
halving, multiplies a process by a positive dyadic rational,
$\frac{n}{2^m} \star p = n \star (p/2^m)$; every process $p$ has an
inverse $-p$ (\cref{prop:inv}), and $\frac{-n}{2^m} \star p = n \star
((-p)/2^m)$.

Tensor products of actions build Pauli strings. On scalars,
$\delta_k \otimes \delta_m = \delta_{k+m}$, with signs multiplying as
for numbers, and $d \otimes \id = \id \otimes d = \id$. On bare actions
$\otimes$ distributes over $\oplus$, $(\bigoplus_i d_i) \otimes
(\bigoplus_j d'_j) = \bigoplus_i \bigoplus_j d_i \otimes d'_j$, and on
actions it tensors the transformations and the bare actions separately,
$uau^\dagger \otimes vbv^\dagger = (u \otimes v)(a \otimes b)(u \otimes
v)^\dagger$, padding a bare action with the identity transformation
$\mathsf{I}$. For example, with $Z = \delta_0 \oplus -\delta_0$ and
$X = \mathsf{H} Z \mathsf{H}^\dagger$,
\[
  Z \otimes X = (\mathsf{I} \otimes \mathsf{H})
  (\delta_0 \oplus -\delta_0 \oplus -\delta_0 \oplus \delta_0)
  (\mathsf{I} \otimes \mathsf{H})^\dagger,
\]
which denotes the Pauli string $Z \otimes X$ (\cref{s:semantics}).
Weighting Pauli strings by dyadic rationals approximates any
Hamiltonian, since the dyadic rationals are dense in the reals: with
$I = \delta_0 \oplus \delta_0$, the two-qubit Hamiltonian
$a\, I \otimes X + b\, X \otimes I + c\, Z \otimes Z$ with real weights
is approximated by $(\frac{n_1}{2^{m_1}} \star I \otimes X) \ppar
(\frac{n_2}{2^{m_2}} \star X \otimes I) \ppar (\frac{n_3}{2^{m_3}} \star
Z \otimes Z)$, with each $\frac{n_i}{2^{m_i}}$ close to its weight.

\subsection{Process equivalence}\label{s:equiv}
When can one process replace another, or part of another, without
changing the behaviour of the program? The answer licenses
optimisation, and rewriting a process into one about which a property
is easier to prove. We give it through \emph{schedules}, which also
define the sequential executions that, halved and repeated, approximate
the concurrent one by the Lie-Trotter formula (\cref{s:semantics}).

\begin{definition}
  A \emph{schedule} is a list of actions $[a_1, \dots, a_m]$. The
  schedule of a process lists its actions, $\sch(a) = [a]$ and
  $\sch(q \ppar r) = \sch(q) \append \sch(r)$, where $\append$
  concatenates lists, and halving acts on schedules element-wise,
  $[a_1, \dots, a_m]/2^n = [a_1/2^n, \dots, a_m/2^n]$.
\end{definition}

Equating schedules needs the sequential behaviour of actions, and
$\Pi_k$ is a complete calculus for sequential quantum programs: \Hamilton
actions are exactly representable as $\Pi_k$ terms by the function
$\pisem{-}$ of \cref{fig:pisem}.
\begin{figure}
  \textbf{Transformations}
  \begin{align*}
    \pisem{\mathsf{I}} & = \mathrm{id}_2 &
    \pisem{\mathsf{H}} & = \zeta_3^\dagger \bullet
    V (\mathrm{id}_1 + \zeta_2) V &
    \pisem{\mathsf{S}} & = \mathrm{id}_1 + \zeta_2 \\
    \pisem{\mathsf{CX}} & = \mathrm{id}_2 + X & &&
    \pisem{\mathsf{SWAP}} & = \sigma_\times \\
    \pisem{u \cdot v} & = \pisem{u} \pisem{v} & &&
    \pisem{u \otimes v} & = \pisem{u} \times \pisem{v}
  \end{align*}

  \vspace{\baselineskip}

  \textbf{Actions}
  \begin{align*}
    \pisem{\id} & = \mathrm{id}_1 &
    \pisem{\delta_k} & = \zeta_k^\dagger &
    \pisem{-\delta_k} & = \zeta_k \\
    \pisem{d \oplus d'} & = \pisem{d} + \pisem{d'} &&&
    \pisem{u a u^\dagger} & = \pisem{u} \pisem{a} \pisem{u}^\dagger
  \end{align*}
  \Description{Equations translating Hamilton into Pi k. Transformations: I
    is the identity on 2; H is zeta 3 dagger times the composite of V, the
    phase id 1 plus zeta 2, and V; S is id 1 plus zeta 2; CX is id 2 plus X;
    SWAP is sigma times; sequential composition becomes composition and
    tensor product becomes product. Actions: id becomes id 1, delta k becomes
    zeta k dagger, minus delta k becomes zeta k, direct sum becomes sum, and a
    conjugation u a u dagger becomes the composite of the images.}
  \caption{The interpretation of \Hamilton transformations and actions
    as terms in the complete calculus of sequential quantum programs
    $\Pi_k$. Terms of $\Pi_k$ compose in sequence by juxtaposition, and
    by direct sum $+$ and tensor product $\times$; $\mathrm{id}_n$ is the
    identity on $n$ elements, and $2 = 1 + 1$ is a qubit; $X$ exchanges
    the two summands of $2$ (the \textsc{not} gate) and $\sigma_\times$
    the two factors of a product, matching the $X$ and $\sigma_\otimes$
    of \citet{freeq};
    $V = \frac12(\begin{smallmatrix} 1+i & 1-i \\ 1-i & 1+i
    \end{smallmatrix})$ is the square root of \textsc{not};
    $\zeta_k = e^{2\pi i/2^k}$ is a scalar, ${\bullet}$ multiplies by a
    scalar, and $(-)^\dagger$ inverts. Since
    $\usem{\delta_k} = e^{-2\pi i/2^k}$, the action $\delta_k$ becomes
    $\zeta_k^\dagger$.}
  \label{fig:pisem}
\end{figure}
With this in hand, we are ready to define schedule equivalence:

\begin{definition}
  Define \emph{schedule equivalence} $\sim$ as the least equivalence
  relation satisfying the following:
  \begin{align}
  [a_1, a_2, \dots, a_m] & \sim [a_{\sigma(1)}, a_{\sigma(2)}, \dots,
    a_{\sigma(m)}] \tag{R1}\label{rule:perm} \\
  [a_1, \dots, a_{j-1}, a_j, a_{j+1}, \dots, a_m]
  & \sim [a_1, \dots, a_{j-1}, b_1, \dots, b_l, a_{j+1}, \dots, a_m]
  \tag{R2} \label{rule:sub}
  \end{align}
  where in \eqref{rule:perm}, $\sigma$ is any permutation of $m$
  elements, and \eqref{rule:sub} can be applied iff there exist
  actions $b_1, \dots, b_l$ and a proof in $\Pi_k$ that
  $\sem{a_j/2^n}_{\Pi} = \sem{b_1/2^n}_{\Pi}\dots \sem{b_l/2^n}_{\Pi}$
  for all $n \in \mathbb{N}$.
\end{definition}

\begin{definition}
  Define \emph{process equivalence} $\approx$ as $p\approx p'$ if and
  only if $\sch(p)\sim \sch(p')$.
\end{definition}

Rule~\eqref{rule:perm} lets the actions of a schedule trade places
freely, which may be surprising for a notion named after an order of
events. A schedule carries two readings of a process. Read additively,
as in the Hamiltonian semantics of \cref{s:semantics}, its actions are
summed and their order is immaterial; this is the reading that process
equivalence captures. Read sequentially, as in the approximate
semantics, its actions form one Trotter step and run in the order
listed; the order then changes the approximation at each finite level,
though never its limit, and \cref{s:implementation} shows by how much.
Swapping two adjacent actions that commute changes nothing even at a
finite level, so the finite levels respect the trace equivalence
of~\citet{mazurkiewicz:traces}, with commutation as independence; rule
\eqref{rule:perm} extends it to all permutations, at the cost of Trotter
error. Equivalence could equally be stated on processes, up to
associativity and commutativity of $\ppar$; we state it on schedules
because a schedule is also the program the compiler runs.

The side condition of rule~\eqref{rule:sub} quantifies over every
halving, and it must. The action $\delta_0 \oplus \id$ turns the phase
of $\ket{1}$ once around the circle, so its unitary is the identity,
like that of $\id \oplus \id$. In parallel with the process
$\mathsf{H}(\delta_2 \oplus -\delta_2)\mathsf{H}$, whose Hamiltonian
is $X/4$, the two differ: started in $\ket{0}$, the composition with
$\id \oplus \id$ ends in $\ket{1}$ with probability \exAIdle, and the
composition with $\delta_0 \oplus \id$ with probability \exATurn. Run at level~$0$ their schedules agree
(probability \exATurnZero{} for both); from level~$2$ on they do not
(\exATurnTwo{} at level $2$), because the other process acts between
the slices of the full turn. \citet{brookes:sharedvariable} met the same
phenomenon in shared-variable parallel programs: \texttt{skip} and
\texttt{x := x+1; x := x-1} have the same input-output behaviour, yet a
concurrent \texttt{y := x} tells them apart, so a compositional
semantics must record behaviour wherever the environment can interfere.
Here the environment interferes at every slice of every level, so
actions may be exchanged only if they agree after every halving, which
is equality of their Hamiltonians, the semantics for which the theory
is complete (\cref{s:completeness}).

We conclude this section by establishing some properties of schedule
and process equivalence, and of \Hamilton processes under process
equivalence. Some of the more straightforward proofs have been relegated
to the supplementary material.

\begin{proposition}
  Process equivalence $\approx$ is an equivalence relation.
\end{proposition}
\begin{proof}
  Immediate, since $\sim$ is an equivalence relation by construction.
\end{proof}

\begin{proposition}\label{prop:cong}
  Process equivalence is congruent under $\ppar$: $p \approx p'$
  implies $p \ppar q \approx p' \ppar q$ and similarly on the right.
\end{proposition}
\begin{proof}
  Suppose that $p \approx p'$, i.e., $\sch(p) \sim \sch(p')$, and let
  $q$ be some other process. By definition $\sch(p \ppar q) = \sch(p)
  \append \sch(q)$.  That $\sch(p) \sim \sch(p')$ means $\sch(p)$ can
  be turned into $\sch(p')$ by a sequence of applications of the
  permutation and/or the substitution rule. But these can be directly
  extended to applications on $\sch(p \ppar q) = \sch(p) \append
  \sch(q)$ only acting on the first part of the schedule: applications
  of the permutation rule can be extended by choosing a permutation
  that does the same on the first part while leaving the other intact,
  and similarly for the substitution rule. Thus $\sch(p \ppar q) \sim
  \sch(p' \ppar q)$, so $p \ppar q \approx p' \ppar q$. The analogous
  argument applies on the other side.
\end{proof}

\begin{proposition}\label{prop:abelian}
  Up to process equivalence, the set of \Hamilton processes of type
  $n$ forms a commutative monoid with $\ppar$ as composition and
  $\id_n$ as identity.
\end{proposition}

\begin{proposition}\label{prop:inv}
  Every process $p : n$ has an inverse process $-p : n$ such that $p
  \ppar (-p) \approx \id_n$.
\end{proposition}

Process equivalence is defined by rules on syntax, but
rule~\eqref{rule:sub} brings in the semantics through $\Pi_k$. Its side
condition asks for a proof in $\Pi_k$, and such a proof exists exactly
when both sides denote the same unitary after every halving, because
the equational theory of $\Pi_k$ is complete~\cite{freeq}. This is
what provides inverses: the inverse $-a$ of an action negates its
scalars, so that $\pisem{(-a)/2^n} = \pisem{a/2^n}^\dagger$ for every
$n$, and rule~\eqref{rule:sub} rewrites the schedule $[\id_n]$ into
$[a, -a]$, as the supplementary material details.

\begin{corollary}
  Up to process equivalence, the set of \Hamilton processes of type $n$
  forms an abelian group with $\ppar$ as composition and $\id_n$ as identity.
\end{corollary}
\begin{proof}
  Combine Propositions \ref{prop:abelian} and \ref{prop:inv}.
\end{proof}

\section{Semantics}\label{s:semantics}

There are several obvious candidates for a semantics of \Hamilton
processes: We can consider the direct Hamiltonian of a process, or the
induced unitary semantics. We can also consider the schedule of a
process as defining a semantics in terms of its sequential
approximations, at both a level $m$ and in the limit as $m \to
\infty$. In fact, we will do all four things, and discuss how these
relate.
\begin{figure}
\textbf{Unitary transformations}
\begin{align*}
\usem{\mathsf{I}} & = \left(\begin{smallmatrix}
  1 & 0 \\ 0 & 1
\end{smallmatrix}\right) &
\usem{\mathsf{H}} & = \tfrac{1}{\sqrt{2}} \left(\begin{smallmatrix}
  1 & 1 \\ 1 & -1
\end{smallmatrix}\right) &
\usem{\mathsf{S}} & = \left(\begin{smallmatrix}
  1 & 0 \\ 0 & i
\end{smallmatrix}\right) \\
\usem{\mathsf{CX}} & = \left(\begin{smallmatrix}
  1 & 0 & 0 & 0 \\ 0 & 1 & 0 & 0 \\ 0 & 0 & 0 & 1 \\ 0 & 0 & 1 & 0
\end{smallmatrix}\right) &
& &
\usem{\mathsf{SWAP}} & = \left(\begin{smallmatrix}
  1 & 0 & 0 & 0 \\ 0 & 0 & 1 & 0 \\ 0 & 1 & 0 & 0 \\ 0 & 0 & 0 & 1
\end{smallmatrix}\right) \\
\usem{u \cdot v} & = \usem{u} \usem{v} &
& &
\usem{u \otimes v} & = \usem{u} \otimes \usem{v}
\end{align*}

\textbf{Hamiltonian actions}
\begin{align*}
  \hsem{\id} & = ( 0 ) & \hsem{\delta_k} & = \left( \tfrac{1}{2^k} \right) &
  \hsem{-\delta_k} & = \left( -\tfrac{1}{2^k} \right) \\
  \hsem{d \oplus d'} & = \hsem{d} \oplus \hsem{d'} & &&
  \hsem{uau^\dagger} & = \usem{u} \hsem{a} \usem{u}^\dagger
\end{align*}

\textbf{Schedules}
\begin{align*}
  \usem{[a_1, \dots, a_k]} = \prod_{j=1}^k \usem{a_j}
\end{align*}

\vspace{0.5\baselineskip}

\textbf{Processes}
\begin{align*}
  \hsem{\ppar_{j=1}^n p_j} & = \sum_{j=1}^n \hsem{p_j} &&&
  \usem{p} & = \exp\left(-2 \pi i \hsem{p} \right) \\
  \snsem{p}{m} & = \usem{\sch(p)/2^m}^{2^m} &&&
  \ssem{p} & = \lim_{m \to \infty} \snsem{p}{m}
\end{align*}

\Description{Equations defining four semantics. Unitary transformations: the 2
  by 2 matrices of I, H and S and the 4 by 4 matrices of CX and SWAP, with
  sequential composition as matrix product and tensor as Kronecker product.
  Hamiltonian actions: id is 0, delta k is 1 over 2 to the k, minus delta k is
  its negative, direct sums are block diagonal, and a conjugation u a u dagger
  is U A U dagger. A schedule denotes the product of the unitaries of its
  actions. A parallel composition denotes the sum of the Hamiltonians; the
  unitary of a process is the exponential of minus 2 pi i times its
  Hamiltonian; the level m schedule semantics is the 2 to the m-th power of
  the halved schedule; and the limit schedule semantics is its limit as m
  grows.}
\caption{The Hamiltonian semantics $\hsem{-}$, unitary semantics
  $\usem{-}$, approximate schedule semantics $\snsem{-}{m}$ at level
  $m \in \mathbb{N}$, and limit schedule semantics $\ssem{-}$ of processes. On
  matrices $M$ and $N$, $M \oplus N$ denotes the block diagonal matrix
  which has $M$ and $N$ as diagonal blocks (in that order) and zeros
  everywhere else, while $M \otimes N$ denotes the Kronecker product
  of $M$ and $N$.}
\label{fig:sem}
\end{figure}
We define the unitary semantics $\usem{-}$, the Hamiltonian semantics
$\hsem{-}$, and the approximate schedule semantics $\snsem{-}{m}$ at
level $m \in \mathbb{N}$, and the limit schedule semantics $\ssem{-}$ in
Figure~\ref{fig:sem}. Note that the Hamiltonian semantics depends on
the unitary semantics for defining the meaning of a conjugated action:
this is justified by the fact that $UHU^\dagger$ is Hermitian whenever
$U$ is unitary and $H$ is Hermitian since
$$
(UHU^\dagger)^\dagger = U^{\dagger\dagger} H^\dagger U^\dagger = U H U^\dagger \enspace.
$$
Similarly, the schedule semantics depend on the unitary semantics of actions,
which in turn is derived from the Hamiltonian semantics of actions. For this
reason, it is more convenient to define the three simultaneously.

There are direct relationships between these semantics, and between the
semantics and process equivalence, that we will explore in this section.
One very direct relationship is between the Hamiltonian and unitary semantics:

\begin{proposition}\label{prop:hamunit}
  The Hamiltonian semantics are strictly more fine-grained than the unitary
  semantics: $\hsem{p} = \hsem{q}$ implies $\usem{p} = \usem{q}$ but
  not vice versa.
\end{proposition}
\begin{proof}
  If $\hsem{p} = \hsem{q}$ then by definition $\usem{p} = \exp(-2 \pi
  i \hsem{p}) = \exp(-2 \pi i \hsem{q}) = \usem{q}$. The converse implication
  does not hold since exponentiation mods out eigenvalues by $2\pi$: e.g.,
  $\usem{\delta_0} = \exp(-2 \pi i \hsem{\delta_0}) = \exp(-2 \pi i) = 1 =
  \exp(0) = \usem{\id}$.
\end{proof}

As regards the schedule semantics, we will see in
Proposition~\ref{prop:ssem} that it does, in fact, produce the same
semantics as the unitary semantics in the limit. To this end, we need a
technical lemma.

\begin{lemma}\label{lem:div}
  For all processes $p$, $\hsem{p/2^m} = \hsem{p}/2^m$.
\end{lemma}
\begin{proof}[Proof sketch]
  By straightforward algebraic manipulation.
\end{proof}

\begin{proposition}\label{prop:ssem}
  For all processes $p$, $\ssem{p} = \usem{p}$.
\end{proposition}
\begin{proof}
  By definition a process $p$ is of the form $p = \ppar_{j=1}^n a_j$
  for actions $a_j$ up to a choice of bracketing, but since $\ppar$ is
  fully associative on the nose in its schedule semantics, this choice
  is semantically irrelevant in that $\sch(p)/2^m = [a_1/2^m, a_2/2^m,
    \dots, a_n/2^m]$ regardless. Then
  \begin{align*}
    \ssem{p}
    & = \lim_{m \to \infty} \snsem{\ppar_{j=1}^n a_j}{m} \\
    & = \lim_{m \to \infty} \usem{\sch(p)/2^m}^{2^m} \\
    & = \lim_{m \to \infty} \left(\prod_{j=1}^n \usem{a_j/2^m}\right)^{2^m} \\
    & = \lim_{m \to \infty} \left(\prod_{j=1}^n
    \exp(-2\pi i \hsem{a_j/2^m})\right)^{2^m} \\
    & = \lim_{m \to \infty} \left(\prod_{j=1}^n
    \exp(-2\pi i \hsem{a_j}/2^m)\right)^{2^m} \\
    & = \exp(-2\pi i \sum_{j=1}^n \hsem{a_j}) \\
    & = \exp(-2\pi i \hsem{\ppar_{j=1}^n a_j}) \\
    & = \exp(-2\pi i \hsem{p}) \\
    & = \usem{p}
  \end{align*}
  where the step that removes the limit is the Lie-Trotter product
  formula~\cite{trotter,hall:lie}, which holds for any finite family of
  matrices and so in particular along the subsequence of powers of two.
  The other steps rewrite by definition and by \cref{lem:div}, so the
  argument takes a single limit and never exchanges two.
\end{proof}
\Cref{prop:ssem} is a guarantee about compilation as a whole: every
process compiles to circuits that converge to its unitary. Like type
soundness, it is established once for the language, rather than checked
on each compiled circuit.

The situation is a bit more delicate with the $m$-level schedule
semantics, since these are not exact but approximate semantics. That
is, $\snsem{p}{m}$ does not correspond to the true semantics of $p$,
merely an approximation only given access to subdivisions of $p$ of
size $2^m$. As such, we cannot expect $\ssem{p} = \ssem{q}$ to imply
$\snsem{p}{m} = \snsem{q}{m}$ since the latter could be false simply
due to $\snsem{p}{m}$ and $\snsem{q}{m}$ being different
approximations of the same limit. Accounting for this fact leads to
the following result:

\begin{proposition}\label{prop:limsem}
  The limit schedule semantics are approximately more fine grained
  than the schedule semantics at \emph{any} particular level $m \in
  \mathbb{N}$: $\ssem{p} = \ssem{q}$ implies that
  $\|\snsem{p}{m} - \snsem{q}{m}\| \to 0$ as $m \to \infty$, where
  $\|\cdot\|$ is the operator norm.
\end{proposition}
\begin{proof}[Proof idea]
  Use that $\snsem{p}{m} = \usem{\sch(p)/2^m}^{2^m}$ and $\snsem{q}{m} = \usem{\sch(q)/2^m}^{2^m}$ both converge to $\ssem{p} = \ssem{q}$.
\end{proof}

We can wonder at this stage how the semantics relate to the notion of
process equivalence we defined previously. We show below that process
equivalence is sound with respect to the most fine-grained of these
semantics, namely the Hamiltonian semantics---so also with respect to
the unitary and limit schedule semantics. The converse direction is
the subject of the next section.

\begin{theorem}[Soundness]\label{thm:sound}
  For processes $p$ and $q$, $p \approx q$ implies $\hsem{p} =
  \hsem{q}$.
\end{theorem}
\begin{proof}
  We start by noticing that if $p$ has the schedule $[a_1, \dots, a_k]$
  then $\hsem{p} = \sum_{j=1}^k \hsem{a_j}$. To prove soundness, we
  proceed by induction on the number of applications of
  \eqref{rule:perm} and \eqref{rule:sub} applied. In the base case, no
  rules are applied and $\hsem{p} = \hsem{q}$ directly. In the
  inductive case, we assume that all proofs using $n$ applications of
  \eqref{rule:perm} and \eqref{rule:sub} are sound, and must show that
  all proofs using $n+1$ applications are sound, too. This yields two
  cases depending on which rule is applied. If the final rule is
  \eqref{rule:perm} by some permutation $\sigma$, then $\sum_{j=1}^k
  \hsem{a_j} = \sum_{j=1}^k \hsem{a_{\sigma(j)}}$ since summation is
  commutative. If the final rule is \eqref{rule:sub}, notice first
  that $\exp(-2\pi i \hsem{p}) = \exp(-2\pi i \hsem{q})$ iff $\hsem{p}
  = \hsem{q}$ when $\|\hsem{p}\|, \|\hsem{q}\| < \frac{1}{2}$ (where
  $\| \cdot \|$ denotes the operator norm), since the function $x
  \mapsto \exp(-ix)$ is bijective on the half-open interval
  $[-\pi,\pi)$. Further, we remark that $\hsem{p} = \hsem{q}$ iff
    $\alpha \hsem{p} = \alpha \hsem{q}$ for (real) nonzero $\alpha$,
    so by choosing $m$ to be sufficiently large, we have that
    $\hsem{p/2^m} = \hsem{p}/2^m = \hsem{q}/2^m = \hsem{q/2^m}$ iff
    $\usem{p/2^m} = \usem{q/2^m}$. Suppose now that $\pisem{a/2^m} =
    \pisem{b_1/2^m}\dots\pisem{b_k/2^m}$ for all $m$, and choose $n$
    such that both $\|\hsem{a}\|/2^n < \frac{1}{2}$ and
    $\|\sum_{j=1}^k \hsem{b_j}\|/2^n < \frac{1}{2}$. Then
  \begin{align*}
    \exp(-2\pi i \hsem{a}/2^n)
    & = \lim_{m \to \infty} \exp(-2\pi i \hsem{a}/2^{n}) \\
    & = \lim_{m \to \infty} \exp(-2\pi i \hsem{a/2^{n}}) \\
    & = \lim_{m \to \infty} (\exp(-2\pi i \hsem{a/2^n}/2^m))^{2^m} \\
    & = \lim_{m \to \infty} \left(\exp(-2\pi i \hsem{a/2^{n+m}})\right)^{2^m} \\
    & = \lim_{m \to \infty} \usem{a/2^{n+m}}^{2^m} \\
    & = \lim_{m \to \infty} \left(\prod_{j=1}^k \usem{b_j/2^{n+m}}\right)^{2^m} \\
    & = \lim_{m \to \infty} \left(\prod_{j=1}^k \exp(-2\pi i
    \hsem{b_j/2^{m+n}})\right)^{2^m} \\
    & = \lim_{m \to \infty} \left(\prod_{j=1}^k
    \exp(-2\pi i \hsem{b_j}/2^{m+n})\right)^{2^m} \\
    & = \exp(-2\pi i \left(\sum_{j=1}^k \hsem{b_j}\right)/2^n)
  \end{align*}
  using the fact that $\hsem{p/2^n} = \hsem{p}/2^n$ for all $p$ and
  $n$ by Lemma~\ref{lem:div}, soundness of $\Pi_k$ (so that
  $\pisem{u} = \pisem{v}$ implies $\usem{u} = \usem{v}$), and the
  Lie-Trotter product formula for the final limit. Since both
  $\hsem{a}/2^n$ and $\sum_{j=1}^k \hsem{b_j}/2^n$ have norm below
  $\frac{1}{2}$, it follows that $\hsem{a}/2^n =
  \sum_{j=1}^k \hsem{b_j}/2^n$, so $\hsem{a} = \sum_{j=1}^k
  \hsem{b_j}$.
\end{proof}

We summarise the relationship between the different semantics in
Figure~\ref{fig:granularity}.
\begin{figure}
  \small
  $$
  \colorbox{green!10}{$\hsem{p} = \hsem{q}$}
  \overset{\text{(Prop.~\ref{prop:hamunit}})}{\Rightarrow}
  \colorbox{blue!10}{$\usem{p} = \usem{q}$}
  \overset{\text{(Prop.~\ref{prop:ssem}})}{\Leftrightarrow}
  \colorbox{purple!10}{$\ssem{p} = \ssem{q}$}
  \overset{\text{(Prop.~\ref{prop:limsem}})}{\Rightarrow}
  \colorbox{red!10}{$\|\snsem{p}{m} - \snsem{q}{m}\| \to 0$ \enspace ($m \to \infty$)}
  $$
  \Description{Four boxed statements about processes p and q joined by
    arrows: equal Hamiltonian semantics implies equal unitary semantics, which
    is equivalent to equal limit schedule semantics, which implies that the
    distance between their level m schedule semantics tends to 0 as m tends
    to infinity.}
  \caption{The semantics of \Hamilton and their relationships.}
  \label{fig:granularity}
\end{figure}

\subsection{Universality}
\label{s:schedules}

So far, we have introduced the syntax of \Hamilton, its associated
notion of process equivalence, and its semantics. Before moving on to
full abstraction, we consider the important question of what can
actually be expressed in \Hamilton.

Recall that the Hermitian operators on $\mathbb{C}^m$ form a
\emph{real} vector space $\mathcal{H}(m)$ of dimension
$m^2$. Focusing, as is commonly done, on the qubit case where $m=2^n$,
a convenient basis for $\mathcal{H}(2^n)$ is given by the Pauli
strings. We show that \Hamilton can express \emph{essentially} all
Hamiltonians in $\mathcal{H}(2^n)$, by showing that $\Hamilton$
processes of type $2^n$ are in a semantic one-to-one correspondence
with Hamiltonians in $\mathcal{H}(2^n)$ whose Pauli coefficients fall
in the ring $\mathbb{Z}[\tfrac{1}{2}]$ of dyadic rationals. Since the
dyadic rationals are dense in the real numbers, any real number
coefficient can be approximated to within an arbitrary error by a
dyadic rational, allowing any Hamiltonian in $\mathcal{H}(2^n)$ to be
approximated in this way by a process in \Hamilton. This is entirely
analogous to the usual notion of universality in the context of
quantum circuits, where a gate set is universal if it allows any unitary
to be approximated to with any desired error.

\begin{theorem}\label{thm:characterisation}
  For every process $p : 2^n$, $\hsem{p}$ has Pauli coefficients in
  the ring $\mathbb{Z}[\tfrac{1}{2}]$ of dyadic rationals.
\end{theorem}
\begin{proof}
  Let $p = \ppar_j a_j$ for actions $a_j$, so $\hsem{p} =
  \hsem{\ppar_j a_j} = \sum_j \hsem{a_j}$, and it suffices to show
  that each $a_j$ has Pauli coefficients in the dyadic rationals.

  If $a_j$ is a bare action then $\hsem{a_j}$ is a diagonal matrix
  with entries of the form $\hsem{\id} = 0$, $\hsem{\delta_k} =
  \frac{1}{2^k}$, or $\hsem{-\delta_k} = -\frac{1}{2^k}$, which are all
  dyadic rational.  This diagonal matrix can be expressed as a sum of
  Pauli strings with only $Z$ and $I$ factors using the fact that
  $$
  \frac{1}{2}(Z + I) = \frac{1}{2}\left(\begin{pmatrix}
    1 & 0 \\ 0 & -1
  \end{pmatrix} + \begin{pmatrix}
    1 & 0 \\ 0 & 1
  \end{pmatrix} \right) = \begin{pmatrix}
    1 & 0 \\ 0 & 0
  \end{pmatrix}
  $$
  and
  $$
  \frac{1}{2}(Z - I) = \frac{1}{2}\left(\begin{pmatrix}
    1 & 0 \\ 0 & -1
  \end{pmatrix} - \begin{pmatrix}
    1 & 0 \\ 0 & 1
  \end{pmatrix} \right) = \begin{pmatrix}
    0 & 0 \\ 0 & 1
  \end{pmatrix}
  $$ so all matrix units $\ket{i}\bra{i}$ in the diagonal can be
  constructed by Kronecker products of these. Denoting the
  $i^\mathrm{th}$ component in the diagonal of $\hsem{a_j}$ by
  $\alpha_{j,i}$, we have $\hsem{a_j} = \sum_i \alpha_{j,i}
  \ket{i}\bra{i}$, and since $\alpha_{j,i}$ is dyadic rational, and
  $\ket{i}\bra{i}$ was constructed above as a sum of Pauli strings with
  dyadic rational coefficients, we have shown that $\hsem{a_j}$ has
  dyadic rational Pauli coefficients.

  In the case where $a_j$ is not a bare action so that $a_j = u b_j
  u^\dagger$ for a bare action $b_j$, we have that $\hsem{a_j} =
  \hsem{u b_j u^\dagger} = \usem{u} \hsem{b_j} \usem{u}^\dagger$. In
  this case, we decompose $\hsem{b_j}$ exactly as above, and remark
  that $\usem{u}$ is a Clifford group operation (so conjugates Pauli
  strings to Pauli strings), so $\usem{u} \hsem{b_j} \usem{u}^\dagger$
  is already a sum of Pauli strings with dyadic rational coefficients.
\end{proof}

\begin{theorem}\label{thm:universality}
  For every $2^n \times 2^n$ Hamiltonian $H$ with Pauli coefficients
  in the ring $\mathbb{Z}[\tfrac{1}{2}]$ of dyadic rationals, there
  exists a process $p : 2^n$ such that $\hsem{p} = H$.
\end{theorem}
\begin{proof}
  Expressing $H$ in the Pauli basis $H = \sum_{j=1}^q \alpha_j P_j$, by
  assumption $\alpha_j \in \mathbb{Z}[\tfrac{1}{2}]$ and $P_j$ is some
  Pauli string. By the assumption, we are free to let $\alpha_j =
  \frac{n}{2^m}$ for some $n$ and $m$. Since $P_j$ is Pauli string, it
  follows that it has eigenvalues $\pm 1$, and so $\alpha_j P_j$ has
  eigenvalues $\pm \frac{n}{2^m}$. Since $\hsem{\delta_m} =
  \frac{1}{2^m}$ and $\hsem{-\delta_m} = -\frac{1}{2^m}$, it follows
  that we can represent the diagonal matrix of eigenvalues of $P_j$ as
  \begin{align*}
    \alpha_j P_j & = \sum_i \ket{i}\bra{i} (-1)^{f(i)} \frac{n}{2^m} \\
    & = n \sum_i \ket{i}\bra{i} (-1)^{f(i)} \frac{1}{2^m} \\
    & = n \sum_i \ket{i}\bra{i} \hsem{(-1)^{f(i)} \delta_m} \\
    & = n \hsem{\bigoplus_i (-1)^{f(i)} \delta_m} \\
    & = \sum_{j=1}^n \hsem{\bigoplus_i (-1)^{f(i)} \delta_m} \\
    & = \hsem{\ppar_{j=1}^n \bigoplus_i (-1)^{f(i)} \delta_m}
  \end{align*}
  where $f$ is a binary valued function indicating whether the
  $i^{\mathrm{tm}}$ component is positive or negative, and we use the
  abuse of notation $(-1)^{f(i)} \delta_m$ to mean $\delta_m$ when $f(i)
  = 0$ and $-\delta_m$ when $f(i) = 1$. In the case where $P_j$ is not
  diagonal, it must have $X$ or $Y$ tensor factors: for each such
  factor, we conjugate the \Hamilton term above by $\mathsf{H}$ if it
  is an $X$, and by $\mathsf{H} ; \mathsf{S}$ if it is $Y$. (For
  example if $P_j = IXZ$, we represent the eigenvalue diagonal as
  above and conjugate the resulting process by $\mathsf{I}
  \otimes \mathsf{H} \otimes \mathsf{I}$.)

  Representing each Pauli term $\alpha_j P_j$ in this way with a
  process $p_j$ satisfying $\hsem{p_j} = \alpha_j P_j$, it follows
  that
  $$
  H = \sum \alpha_j P_j = \sum_{j=1}^q \hsem{p_j} = \hsem{\ppar_{j=1}^q p_j}
  $$
  as desired.
\end{proof}
In principle, combining these two theorems yields a procedure for
normalisation by evaluation. However, doing so is not always efficient
since $\mathcal{H}(2^n)$ has dimension $4^n$, so normalising a process
in this way may result in an exponentially large term. For processes
whose actions are Pauli strings, the Pauli representation has one string
per action, and \cref{s:implementation} computes it without building
any matrix.

\section{Full abstraction}\label{s:completeness}

The goal of this section is to show the converse of
Theorem~\ref{thm:sound} to obtain a full abstraction result of process
equivalence with regards to the Hamiltonian semantics, i.e., that $p
\approx q$ iff $\hsem{p} = \hsem{q}$ for all processes $p,q$ of type
$2^n$.

The overall strategy is to use the normalisation procedure given by
combining Theorems~\ref{thm:characterisation} and
\ref{thm:universality} to obtain a \emph{Pauli representation} of a
process, and show that every process is equivalent to its Pauli
representation. This is done in several stages.
\begin{enumerate}
  \item First, we consider the case of \emph{simple processes} which
    are those that are free of unitary transformations. These are
    shown to have diagonal Hamiltonian semantics
    (Lemma~\ref{lem:simplediag}), and to normalise under $\approx$ to
    a form that (essentially) normalises each entry in the diagonal
    one-by-one. A consequence of this is that any simple process is
    equivalent to its Pauli representation
    (Corollary~\ref{corr:paulirep}).
  \item Extending this to arbitrary processes relies on showing a few
    tricks: that this normalisation extends under conjugation by a
    unitary (Lemma~\ref{lem:conjcong}), that the action of conjugating
    a Pauli string by a unitary can be normalised
    (Lemma~\ref{lem:paulinorm}), and that the only unitaries permitted
    to transform by are Clifford operations, which conjugate Pauli
    strings to Pauli strings. Together, these are used to show
    (Theorem~\ref{thm:paulirep}) that processes normalise to their
    Pauli representations, and the desired result follows.
\end{enumerate}

\begin{definition}
Say that a process $p$ is \emph{simple} if $p = \ppar_j a_j$ with each
$a_j$ a bare action.
\end{definition}

In other words, a simple process is a process that is free of transformations.

\begin{lemma}\label{lem:simplediag}
  If $p : n$ is a simple process then $\hsem{p}$ is a diagonal matrix.
\end{lemma}
\begin{proof}
  Suppose $p = \ppar_j a_j$ where each $a_j$ is a bare action. Since
  each $a_j$ is a bare action, it follows that $a_j = \bigoplus_{k=1}^n
  s_{j,k}$ where $s_{j,k}$ is either some $\delta_m$, $-\delta_m$, or $\id$.
  As such,
  $$
  \hsem{a_j} = \hsem{\bigoplus_{k=1}^n s_{j,k}} = \bigoplus_{k=1}^n \hsem{s_{j,k}}
  $$
  so $\hsem{a_j}$ is diagonal. But then
  $$
  \hsem{p} = \hsem{\ppar_j a_j} = \sum_j \hsem{a_j} =
  \sum_j \bigoplus_{k=1}^n \hsem{s_{j,k}}
  $$
  so $\hsem{p}$ is a sum of diagonal matrices, i.e., diagonal.
\end{proof}

In this development, we will need some straightforward lemmas (whose
proofs have be deferred to the supplementary material).

\begin{lemma}\label{lem:act}
  For all bare actions $a : n$ and $b : m$, it is the case that
  \begin{enumerate}[(i)]
    \item $a \oplus b \approx a \oplus \id_m \ppar \id_n \oplus b$,\label{lem:act1}
    \item $\id_n \oplus \id_m \approx \id_{n+m}$,\label{lem:act2}
    \item $u(a \oplus b)u^\dagger \approx u(a \oplus \id_m)u^\dagger
      \ppar u(\id_n \oplus b)u^\dagger$,\label{lem:act3}
    \item $u \id_n u^\dagger \approx \id_n$.\label{lem:act4}
  \end{enumerate}
\end{lemma}

\begin{lemma}\label{lem:zetadd}
  For all $k, l, n \in \mathbb{N}$,
  \begin{enumerate}[(i)]
  \item $\id_l \oplus \delta_{k+1} \oplus \id_n \ppar \id_l \oplus \delta_{k+1} \oplus \id_n \approx \id_l \oplus \delta_k \oplus \id_n$ and \label{lem:zetaddpos}
  \item $\id_l \oplus -\delta_{k+1} \oplus \id_n \ppar \id_l \oplus -\delta_{k+1} \oplus \id_n \approx \id_l \oplus -\delta_k \oplus \id_n$. \label{lem:zetaddneg}
  \end{enumerate}
\end{lemma}

\begin{definition}
  Say that a simple process $p = \ppar_j a_j$ is a \emph{$1$-level process}
  if there exist $n,m \in \mathbb{N}$ such that each $a_j = \id_n
  \oplus b_j \id_m$ for $b_j$ of the form $0$, $\delta_k$, or
  $-\delta_k$ for some $k$.
\end{definition}

\begin{lemma}\label{lem:1levelnorm}
  If $p$ is a $1$-level process then either $p \approx \id$ or $p
  \approx \pm \frac{a}{2^b} \star \id_n \oplus \delta_0 \oplus \id_m$
  for unique $a,b,n,m \in \mathbb{Z}_+$ such that $a$ is odd.
\end{lemma}
\begin{proof}
  Since $p = \ppar_j \id_n \oplus b_j \oplus \id_m$ for $b_j$ scalars,
  and since $\ppar$ is unital with $\id$ as unit by
  Proposition~\ref{prop:abelian}, we are free to assume that either $p
  \approx \id$ (in case all $b_j = \id$) or no $b_j = \id$. If the
  former is the case, we are done. If the latter is the case, all
  $b_j$ are of the form $\pm \delta_k$. In this case, we use the
  following procedure to normalise $p$:
  \begin{enumerate}[(i)]
  \item Find the largest $k$ such that $\id_n \oplus \delta_k \oplus
    \id_m$ or $\id_n \oplus (-\delta_k) \oplus \id_m$ appears in $p$,
    and use Lemma~\ref{lem:zetadd} to express every action of the form
    $\id_n \oplus \delta_{k'} \oplus \id_m$ as a composition of a
    number of $\id_n \oplus \delta_k \oplus \id_m$, and every action of
    the form $\id_n \oplus (-\delta_{k'}) \oplus \id_m$ as a
    composition of a number of $\id_n \oplus (-\delta_k) \oplus \id_m$.
  \item If both $\id_n \oplus \delta_k \oplus \id_m$ and $\id_n \oplus
    (-\delta_k) \oplus \id_m$ appear in this expansion, pair up each
    $\id_n \oplus \delta_k \oplus \id_m$ with a $\id_n \oplus
    (-\delta_k) \oplus \id_m$ until the remaining terms are either all
    $\id_n \oplus \delta_k \oplus \id_m$ or all $\id_n \oplus
    (-\delta_k) \oplus \id_m$, then use Proposition~\ref{prop:abelian}
    to get rid of all matching pairs.
  \item At this point, the expansion is either $\id$ or consists of
    either all $\id_n \oplus \delta_k \oplus \id_m$ or all $\id_n
    \oplus (-\delta_k) \oplus \id_m$. If the expansion is $\id$ or the
    number of terms is odd, we are done. If the number of terms is
    even, match up terms pairwise and apply Lemma~\ref{lem:zetadd} to
    combine terms, dividing the number of terms by $2$, and continue
    this process until the number of terms is odd.
  \end{enumerate}
  Since $a/2^b \star \id_n \oplus \delta_0 \oplus \id_m =
  \underbrace{\id_n \oplus \delta_b \oplus \id_m \ppar \cdots \ppar
    \id_n \oplus \delta_b \oplus \id_m}_{a~\text{times}}$ and
  analogously for $-a/2^b \star \id_n \oplus \delta_0 \oplus \id_m$,
  after normalisation the process is on the required form. To see that
  this is unique, we notice that
  \begin{align*}
      \hsem{a/2^b \star \id_n \oplus \delta_0 \oplus \id_m}
  & = \hsem{\underbrace{\id_n \oplus \delta_b \oplus \id_m \ppar \cdots
      \ppar \id_n \oplus \delta_b \oplus \id_m}_{a~\text{times}}} \\
  & = a \hsem{\id_n \oplus \delta_b \oplus \id_m} \\
  & = a \frac{1}{2^b} \ket{n}\bra{n}
  = \frac{a}{2^b} \ket{n}\bra{n}
  \end{align*}
  which is a normal representation of a dyadic rational (as a single
  entry in a 1-level diagonal matrix) since $a$ is odd.
\end{proof}

\begin{proposition}\label{prop:simple1level}
  Every simple process is equivalent to a composition of $1$-level
  processes of the form $\pm \frac{a_j}{2^{b_j}} \star \id_{n_j}
  \oplus \delta_0 \oplus \id_{m_j}$ for unique $a_j,b_j,n_j,m_j$ where
  $a_j$ is odd.
\end{proposition}
\begin{proof}
  Let $p = \ppar_{j=1}^J c_j$ where each $c_j$ is a bare action, i.e., of
  the form $c_j = \bigoplus_l d_{j,l}$ for scalars $d_{j,l}$. Applying
  Lemma~\ref{lem:act} on each $c_j$ yields that $c_j$ is a composition
  of $1$-level processes, namely
  $$
  c_j \approx d_{j,1} \oplus \id_{n-1} \ppar \id_1 \oplus d_{j,2}
  \oplus \id_{n-2} \ppar \cdots \ppar \id_{n-1} \oplus d_{j,l}
  $$
  Grouping the resulting $1$-level processes by index using
  associativity and commutativity (Proposition~\ref{prop:abelian}), we
  get that $p$ is equivalent to a big parallel composition of
  $1$-level processes acting on each component,
  \begin{align*}
      p & \approx (d_{1,1} \oplus \id_{n-1} \ppar \cdots \ppar d_{J,1}
      \oplus \id_{n-1}) \ppar \\
      & \qquad (\id_1 \oplus d_{1,2} \oplus \id_{n-1}
      \ppar \cdots \ppar
      \id_1 \oplus d_{J,2} \oplus \id_{n-1}) \ppar \\
      & \qquad \cdots \ppar \\
      & \qquad (\id_{n-1} \oplus d_{1,n} \ppar \cdots \ppar
      \id_{n-1} \oplus d_{J,n})
  \end{align*}
  and applying Lemma~\ref{lem:1levelnorm} to each group yields the
  desired result.
\end{proof}

\begin{definition}
  For a process $p$, call the process $q$ extracted from $\hsem{p}$ in
  Theorem~\ref{thm:universality} the \emph{Pauli representation} of $p$.
\end{definition}

\begin{corollary}\label{corr:paulirep}
  Every simple process is equivalent to its Pauli representation.
\end{corollary}
\begin{proof}
  Simple processes have diagonal Hamiltonian semantics by
  Lemma~\ref{lem:simplediag}, so their Pauli representation is simple.
  Normalising the process $p$ to a composition of $1$-level processes
  $r$ (using Proposition~\ref{prop:simple1level}) yields $p \approx
  r$, and doing the same for its Pauli representation $q$ yields $q
  \approx r$, so $p \approx q$ since $\approx$ is an equivalence
  relation.
\end{proof}

\begin{corollary}
  For simple processes $p$ and $q$, $\hsem{p} = \hsem{q}$ implies $p
  \approx q$.
\end{corollary}
\begin{proof}
  Since $\hsem{p} = \hsem{q}$ it follows that $p$ and $q$ have the same
  Pauli representation $r$, and since $p \approx r$ and $q \approx r$ by
  Corollary~\ref{corr:paulirep}, it follows that $p \approx q$ by $\approx$
  an equivalence relation.
\end{proof}

\begin{lemma}\label{lem:conjcong}
  $p \approx \ppar_j a_j$ implies $upu^\dagger \approx \ppar_j
  ua_ju^\dagger$ when $p$ and all $a_j$ are bare actions.
\end{lemma}
\begin{proof}
  We proceed by induction on the number $n$ of rule applications in
  the proof that $p \approx \ppar_j a_j$. In the base case $n=0$, so
  no rules are applied, and we are done. In the inductive case, we
  assume that $p \approx \ppar_j a_j$ implies $upu^\dagger \approx
  \ppar_j ua_ju^\dagger$ when the proof requires $n$ rule applications, and
  must show that the same is the case for $n+1$ rule applications.

  There are two cases depending on the final rule applied. When the
  final rule is the permutation rule, we have $p \approx \ppar_j a'_j$
  in $n$ steps, and $\ppar_j a'_j = \ppar_j a_{\sigma(j)}$ for some
  permutation $\sigma$. By induction $upu^\dagger \approx \ppar_j u
  a'_j u^\dagger$, so also $\ppar_j u a'_j u^\dagger = \ppar_j u
  a_{\sigma(j)} u^\dagger$. As such, applying the permutation rule
  with $\sigma$ yields $p \approx \ppar_j u a_{\sigma(j)} u^\dagger$.

  When the final rule applied is the substitution rule, we have $p
  \approx \ppar_j a'_j$ in $n$ steps, so we must have that
  $\sch(\ppar_j a'_j) = [a'_1, \dots, a'_q]$ and $$\sch(\ppar_j a_j) =
  [a'_1, \dots, a'_{k-1}, b_1, \dots, b_l, a'_{k+1}, \dots, a'_q]$$
  where $\pisem{a'_k/2^m} = \prod_{i=1}^l \pisem{b_i/2^m}$. But then
  \begin{align*}
    \pisem{(u a'_k u^\dagger)/2^m}
    & = \pisem{u (a'_k/2^m) u^\dagger} \\
    & = \pisem{u}\pisem{a'_k/2^m}\pisem{u}^\dagger \\
    & = \pisem{u} \left(\prod_{i=1}^l \pisem{b_i/2^m}\right)
    \pisem{u}^\dagger \\
    & \overset{(*)}{=} \prod_{i=1}^l \pisem{u} \pisem{b_i/2^m}
    \pisem{u}^\dagger \\
    & = \prod_{i=1}^l \pisem{u (b_i/2^m) u^\dagger} \\
    & = \prod_{i=1}^l \pisem{(u b_i u^\dagger)/2^m}
  \end{align*}
  where the step marked $(*)$ follows since $\pisem{u}^\dagger
  \pisem{u} = \pisem{u} \pisem{u}^\dagger = \mathrm{id}$. Since we
  must then have $\sch(\ppar_j u a'_j u^\dagger) = [u a'_1 u^\dagger,
    \dots, u a'_q u^\dagger]$ and $$\sch(\ppar_j u a_j u^\dagger) = [u
    a'_1 u^\dagger, \dots, u a'_{k-1} u^\dagger, u b_1 u^\dagger,
    \dots, u b_l u^\dagger, u a'_{k+1} u^\dagger, \dots, u a'_q
    u^\dagger],$$ that $u p u^\dagger \approx \ppar_j u a_j u^\dagger$
  follows by the substitution rule using the proof that
  $\pisem{(u a'_k u^\dagger)/2^m} = \prod_{i=1}^l \pisem{(u b_i
    u^\dagger)/2^m}$.
\end{proof}

\begin{lemma}\label{lem:compadj}
  For all transformations $u,v : n$ and action $a : n$, it is the case that
  $u(vav^\dagger)u^\dagger \approx (u \cdot v) a (u \cdot v)^\dagger$.
\end{lemma}
\begin{proof}
  For any action $a$ (in particular any action of the form $b/2^m$) we have
  \begin{align*}
    \pisem{u(vav^\dagger)u^\dagger}
    & = \pisem{u} \pisem{vav^\dagger} \pisem{u}^\dagger \\
    & = \pisem{u} \pisem{v} \pisem{a} \pisem{v}^\dagger \pisem{u}^\dagger \\
    & = (\pisem{u} \pisem{v}) \pisem{a} (\pisem{u} \pisem{v})^\dagger \\
    & = \pisem{u \cdot v} \pisem{a} \pisem{u \cdot v}^\dagger \\
    & = \pisem{(u \cdot v) a (u \cdot v)^\dagger}
  \end{align*}
  so the statement follows by application of the substitution rule.
\end{proof}

\begin{lemma}\label{lem:paulinorm}
  When $p : 2^n$ is a diagonal Pauli representation and $u$ is any
  transformation, there exists an action $q : 2^n$ such that $u p
  u^\dagger \approx vqv^\dagger$ where $vqv^\dagger$ is a Pauli
  representation.
\end{lemma}
\begin{proof}
  When $p$ is a diagonal Pauli representation, it follows that it
  represents a Pauli string with only $Z$ and $I$ as tensor factors.
  Since $\usem{u}$ is a Clifford operation, it follows that
  $\hsem{upu^\dagger} = \usem{u}\hsem{p}\usem{u}^\dagger$ is again a
  Pauli string, and by Theorem~\ref{thm:universality} that there
  exists some Pauli representation $vqv^\dagger$ such that
  $\hsem{upu^\dagger} = \hsem{vqv^\dagger}$. But then it follows that
  $$\hsem{upu^\dagger/2^m} = \hsem{upu^\dagger}/2^m =
  \hsem{vqv^\dagger}/2^m = \hsem{vqv^\dagger/2^m}$$ for all $m$, and
  in turn that $\usem{upu^\dagger/2^m} = \usem{vqv^\dagger/2^m}$ for
  all $m$. Since $\Pi_k$ is complete, it then follows that
  $\pisem{upu^\dagger/2^m} = \pisem{vqv^\dagger/2^m}$, so the statement
  follows by a single application of the substitution rule.
\end{proof}

\begin{theorem}\label{thm:paulirep}
  Every process $p : 2^n$ is equivalent to its Pauli representation.
\end{theorem}
\begin{proof}
  Let $p = \ppar_j u_j a_j u_j^\dagger$ be a process. Without loss of
  generality we assume that each $a_j$ is a bare action, since if it
  isn't, we can always use Lemma~\ref{lem:compadj} to combine a
  multiply conjugated action into a bare action conjugated by a
  composition of transformations. Similarly, if some term in $p$ is not
  conjugated, we can always see it as conjugated by $\mathsf{I}$.

  Applying Corollary~\ref{corr:paulirep}, Lemma~\ref{lem:conjcong},
  and congruence of process equivalence under $\ppar$
  (Proposition~\ref{prop:cong}) on each $u_j a_j u_j^\dagger$ yields
  that $$p \approx \ppar_i u_i \left(\frac{n_i}{2^{m_i}} \star
  b'_i\right) u_i^\dagger$$ where each $b_i'$ is a representation of a
  diagonal Pauli string, and $u_i$ is some transformation. However,
  since the permitted transformations $u_i$ are precisely the Clifford
  group operations, it follows that each $\hsem{u_i
    \left(\frac{n_i}{2^{m_i}} \star b'_i\right) u_i^\dagger}$ is a
  Pauli string (as Clifford operations conjugate Pauli strings to
  Pauli strings by definition). However, it may not yet be a Pauli
  \emph{representation} as extracted by
  Theorem~\ref{thm:universality}, since there are many non-canonical
  ways of representing each Pauli string. However, using
  Lemma~\ref{lem:paulinorm} we can always rewrite $u_i b'_i
  u_i^\dagger$ in $u_i \left(\frac{n_i}{2^{m_i}} \star
  b'_i\right)u_i^\dagger = \frac{n_i}{2^{m_i}} \star u_i b'_i
  u_i^\dagger$ into its Pauli representation.

  In this way, $p \approx \ppar_i \frac{n_i}{2^{m_i}} \star p_i$
  where each $p_i$ is a Pauli representation. Using associativity and
  commutativity of $\ppar$ (Proposition~\ref{prop:abelian}) to group
  all Pauli strings of the same type, we finally use the same
  technique as in Lemma~\ref{lem:1levelnorm} to combine the
  coefficients $\frac{n_i}{2^{m_i}}$ into one for each type of Pauli
  string, and we are done.
\end{proof}

\begin{corollary}[Completeness]\label{corr:comp}
  For all processes $p : 2^n$ and $q : 2^n$, $\hsem{p} = \hsem{q}$
  implies $p \approx q$.
\end{corollary}
\begin{proof}
  Since $\hsem{p} = \hsem{q}$ it follows that $p$ and $q$ have the same
  Pauli representation $r$, and since $p \approx r$ and $q \approx r$ by
  Theorem~\ref{thm:paulirep}, it follows that $p \approx q$ by $\approx$
  an equivalence relation.
\end{proof}

\begin{theorem}[Full abstraction]
  For all processes $p : 2^n$ and $q : 2^n$, $\hsem{p} = \hsem{q}$ iff
  $p \approx q$.
\end{theorem}
\begin{proof}
  Combine Corollary~\ref{corr:comp} and Theorem~\ref{thm:sound}.
\end{proof}

Full abstraction is what a compiler needs from the theory. Equivalent
programs denote the same Hamiltonian but may compile to circuits that
differ in error and size, so a compiler may replace a program, or any
part of one, by a cheaper equivalent, and check the replacement by
comparing Hamiltonians, which \cref{s:implementation} does by normal
forms without building any matrix.

\subsection{Lie structure}\label{s:lie}
Full abstraction makes the correspondence with Lie algebras exact.
Write $\mathcal{L}(2^n)$ for the set of matrices $-iH$ with $H$ a
$2^n \times 2^n$ Hamiltonian whose Pauli coefficients are dyadic, and
$[A,B] = AB - BA$ for the commutator. Like $\star$, define a derived
form: the \emph{bracket} $[p,q]$ of processes $p, q : 2^n$ is any
process with $\hsem{[p,q]} = -i[\hsem{p}, \hsem{q}]$.

\begin{proposition}\label{prop:lie}
  (i) $\mathcal{L}(2^n)$ is closed under addition, negation and the
  commutator, so it is a Lie algebra over $\mathbb{Z}[\tfrac12]$, and
  its real span is the Lie algebra $\mathfrak{u}(2^n)$ of $U(2^n)$.
  (ii) The bracket of any two processes exists and is unique up to
  $\approx$. (iii) The map $p \mapsto -i\hsem{p}$ is an isomorphism of
  Lie algebras from processes of type $2^n$ up to $\approx$, with
  $\ppar$ as addition and $[-,-]$ as bracket, onto $\mathcal{L}(2^n)$.
\end{proposition}
\begin{proof}
  (i) Two Pauli strings $P$ and $Q$ either commute or anticommute. If
  they anticommute, $[P,Q] = 2PQ = \pm 2iR$ for a Pauli string $R$, so
  $[-i\alpha P, -i\beta Q] = -i(\pm 2\alpha\beta)R$, with dyadic
  coefficient when $\alpha$ and $\beta$ are dyadic; the commutator is
  bilinear, so closure follows. Pauli strings span all matrices, so the
  real span of $\mathcal{L}(2^n)$ is the space of matrices $-iH$ with
  $H$ Hermitian, which is $\mathfrak{u}(2^n)$~\cite{hall:lie}.
  (ii) By (i), $-i[\hsem{p}, \hsem{q}]$ is Hermitian with dyadic Pauli
  coefficients, so a process denoting it exists by
  \cref{thm:universality}, and any two such processes are equivalent by
  full abstraction. (iii) The map is well defined and injective by full
  abstraction and surjective by \cref{thm:universality}; it takes
  $\ppar$ to addition by \cref{fig:sem}; and
  $-i\hsem{[p,q]} = -[\hsem{p}, \hsem{q}] = [-i\hsem{p}, -i\hsem{q}]$.
\end{proof}

The exponential carries $\mathcal{L}(2^n)$ into $U(2^n)$, but in general
outside the unitaries that $\Pi_k$ terms denote: these have algebraic
entries and so algebraic eigenvalues, while $e^{-2\pi i (X+Z)/8}$ has
eigenvalues $e^{\mp i\pi\sqrt2/4}$, which are transcendental by the
Gelfond-Schneider theorem. The schedule semantics replaces the exponential
by $\Pi_k$ circuits that converge to it (\cref{prop:ssem}), and the bracket
measures what that costs: one step $e^{A}e^{B}$ differs from $e^{A+B}$ by
$\tfrac12[A,B]$ to first order, and actions whose bracket is $0$ are the
independent actions of \cref{s:processes}, whose order no level can see.

\section{Implementation}\label{s:implementation}

The semantics of the preceding sections is realised as a working compiler: a
Haskell core that delegates linear algebra and gate synthesis to
numpi and qiskit (both Python libraries). The
pipeline takes a process term, forms the canonical schedule at a chosen
refinement level, lowers it to a quantum circuit, and compares the result
against the ideal continuous evolution for validation. The next three
examples each isolate one characteristic behavior: Trotter convergence,
schedule equivalence, and exact compilation of a commuting interaction. A
larger, physically motivated instance follows in \cref{s:applications}.

A process term is represented directly as the abstract syntax of
\cref{s:processes}:
\begin{minted}{haskell}
data Proc
  = Atom Action
  | Proc :∥: Proc
\end{minted}
\noindent
The compiler entry point is
\mintinline{haskell}{compileTrotter n p},
which forms the canonical schedule at refinement level $n$ (repeated $2^n$ times):
\begin{minted}{haskell}
compileTrotter :: Natural -> Proc -> QCircuit
compileTrotter n proc =
  mconcat (replicate reps step)
  where
    step = compileSchedule (canonicalSchedule proc n)
    reps = 2 ^ (fromIntegral n :: Int)
\end{minted}
\noindent
Each atomic action is a diagonal operator, possibly conjugated by a Clifford.
The scalar generator $\delta_k^{\,j}$ compiles to a phase of $-2\pi j/2^{k}$:
\begin{minted}{haskell}
diagonalAngles :: BareAction -> [Double]
diagonalAngles = \case
  Id             -> [0]
  Delta k        -> [phaseAngle k]
  NegDelta k     -> [-(phaseAngle k)]
  DeltaPow k j   -> [fromIntegral j * phaseAngle k]
  a :⊕: b        -> diagonalAngles a ++ diagonalAngles b
\end{minted}
\noindent
Although the formal syntax of \cref{s:processes} treats $\delta_k^{\,j}$ as the
derived abbreviation $j \star \delta_k$, the implementation promotes it to a
primitive constructor \mintinline{haskell}{DeltaPow k j}. This keeps schedules
compact: expanding a coefficient $j$ into $j$ parallel copies of $\delta_k$ before
compilation would blow up schedule size on Hamiltonians with large integer
coefficients.
\noindent
Diagonal operators are lowered to elementary gates through Qiskit's
\texttt{DiagonalGate}; conjugations become the corresponding basis-change gates
surrounding the diagonal. For validation,
\mintinline{haskell}{compareToIdeal} builds the exact unitary
$e^{-2\pi i\,\llbracket p\rrbracket_H}$ and reports the Frobenius distance to the
compiled circuit, aligned up to global phase.

The building blocks for single-axis rotations are:
\begin{minted}{haskell}
zK :: Natural -> Proc
zK k = act (Id ⊕ Delta k)

xK :: Natural -> Proc
xK k = conj H (Id ⊕ Delta k)

yK :: Natural -> Proc
yK k = conj (S · H) (Id ⊕ Delta k)
\end{minted}
\noindent
and the $X+Z$ Hamiltonian of \cref{s:background}, physically a single spin in
a magnetic field with components along the $z$ and $x$ axes, is then:
\begin{minted}{haskell}
zxK :: Natural -> Natural -> Proc
zxK kz kx = xK kx ∥ zK kz
\end{minted}

Compiling \mintinline{haskell}{zxK} at increasing refinement levels and comparing
against the exact evolution exhibits the central approximation behavior. At level
$n$ the schedule alternates a $Z$-slice and an $X$-slice $2^{n}$ times, each slice
carrying an angle scaled by $2^{-n}$, and the circuit's distance to the ideal
falls accordingly:

\begin{center}
\begin{tabular}{r r r}
\toprule
$n$ & reps $2^n$ & Frobenius \\
\midrule
0 & 1 & 0.4002 \\
1 & 2 & 0.1933 \\
2 & 4 & 0.0958 \\
3 & 8 & 0.0478 \\
4 & 16 & 0.0239 \\
\bottomrule
\end{tabular}

\end{center}

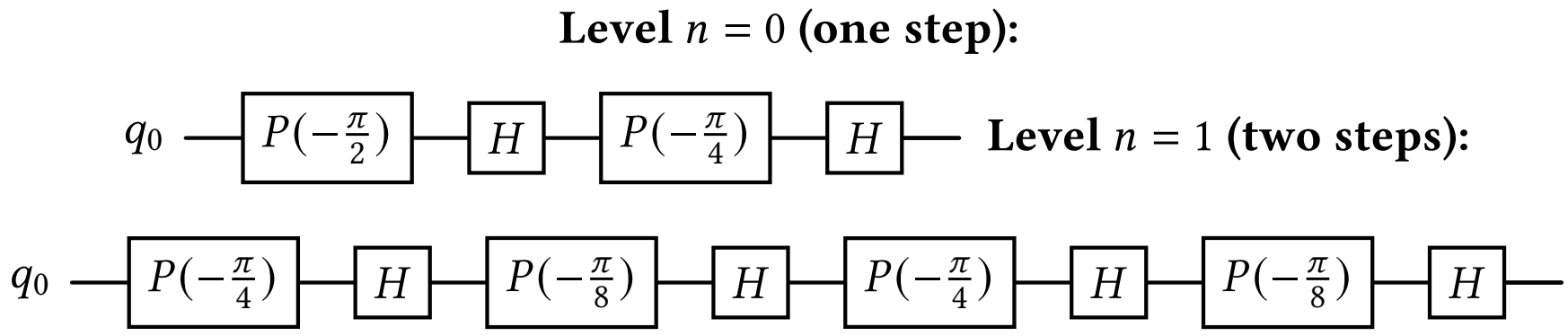
\begin{figure}
\centering
\textbf{Level $n=0$ (one step):}\\[4pt]
\begin{quantikz}[column sep=small,row sep={24pt,between origins}]
\lstick{$q_{0}$} & \gate{P({-}\frac{\pi}{2})} & \gate{H} & \gate{P({-}\frac{\pi}{4})} & \gate{H} & \qw
\end{quantikz}
\textbf{Level $n=1$ (two steps):}\\[4pt]
\begin{quantikz}[column sep=small,row sep={24pt,between origins}]
\lstick{$q_{0}$} & \gate{P({-}\frac{\pi}{4})} & \gate{H} & \gate{P({-}\frac{\pi}{8})} & \gate{H} & \gate{P({-}\frac{\pi}{4})} & \gate{H} & \gate{P({-}\frac{\pi}{8})} & \gate{H} & \qw
\end{quantikz}
\Description{Two single-qubit circuits for zxK 2 3. Level n equal to 0: a
  phase gate P of minus pi over 2, a Hadamard, a phase gate P of minus pi over
  4 and a Hadamard. Level n equal to 1: the same four gates with the phase
  angles halved, to minus pi over 4 and minus pi over 8, repeated twice.}
\caption{Compiled circuits for \texttt{zxK 2 3} at refinement levels $n=0$ and
  $n=1$. $P(\theta)$ denotes the single-qubit phase gate $\mathrm{diag}(1,
  e^{i\theta})$. Increasing the refinement level halves the angles and doubles
  the repetitions, which halves the distance from the ideal evolution at the
  endpoint; \cref{fig:chem} shows a refined circuit tracking the path as
  well.}
\label{fig:circuits}
\end{figure}

\noindent
The error halves with each level, the expected first-order Trotter rate.
\Cref{fig:circuits} shows the compiled circuit at levels $n=0$ and $n=1$:
at level~$0$ a single coarse $Z$-slice followed by an $X$-slice; at level~$1$
two repetitions of half-angle slices, which halve the distance to the ideal.
The process term is unchanged across both; only the refinement level varies.
The distance compares endpoints, the evolutions after one unit of time;
\cref{fig:chem} in \cref{s:applications} shows how closely a refined circuit
also follows the path in between.

Because $\parallel$ commits to no ordering, the canonical schedule is only one of
several admissible realisations, and the implementation enumerates the others by
permuting the schedule and scoring each against the ideal. Two cases are
instructive. For the two-term \mintinline{haskell}{zxK} the step contains a
$Z$-slice and an $X$-slice, which may be applied in either order:
\begin{center}
\begin{tabular}{l r}
\toprule
ordering & Frobenius \\
\midrule
$[Z, X]$ & 0.1933 \\
$[X, Z]$ & 0.1933 \\
\bottomrule
\end{tabular}

\end{center}
The two circuits are genuinely different, yet they are equidistant from the
ideal, and the reason is a symmetry of this example: $Z$ and $X$ are real
symmetric matrices, so reversing the order of the two slices transposes the
unitary of each step, and hence of the whole circuit, while the exact evolution
is itself symmetric, and transposition preserves the distance.

The orderings are always equivalent as realisations of the Hamiltonian, but
without such a symmetry they need not be equally good \emph{approximations}.
Take a three-term single-qubit Hamiltonian, physically a spin in a field with
components along all three axes:
\begin{minted}{haskell}
zxyK :: Proc
zxyK = zK 1 ∥ xK 2 ∥ yK 3
\end{minted}
A dominant $Z$ at $k=1$, a medium $X$ at $k=2$, a small $Y$ at $k=3$, no two of
which commute, and the $Y$ term breaks the symmetry above: $Y$ is antisymmetric,
so transposition reverses its rotation. Enumerating its six orderings at level
$1$ produces a clear ranking:
\begin{center}
\begin{tabular}{l r}
\toprule
ordering & Frobenius \\
\midrule
$[X, Z, Y]$ & 0.4370 \\
$[Y, X, Z]$ & 0.5276 \\
$[Z, Y, X]$ & 0.5472 \\
$[X, Y, Z]$ & 0.6355 \\
$[Z, X, Y]$ & 0.6547 \\
$[Y, Z, X]$ & 0.7292 \\
\bottomrule
\end{tabular}

\end{center}
The best and worst admissible circuits differ in accuracy by a factor of about
$1.7$, and the ranking is structured: the three best orderings are cyclic
rotations of one another, and the three worst are the rotations of the reverse
cycle. Every one of these circuits realizes
the same Hamiltonian; the choice among them is pure optimisation within an
equivalence class the theory certifies, and Trotter error plays the part
that running time plays for classical programs. The equivalence makes any reordering
safe; the quality spread makes the choice matter. A compiler can therefore
search among reorderings for the most accurate one without re-checking what a
candidate computes: soundness (\cref{thm:sound}) guarantees in advance that
every reordering denotes the same Hamiltonian, so the search estimates only how
well each candidate approximates it, and completeness (\cref{corr:comp})
guarantees that the same rules reach every other program with that
Hamiltonian, beyond its reorderings.

Rewriting by the theory also reaches circuits that no circuit optimizer
can. Any rewriting that preserves the unitary of a circuit preserves its
Trotter error; the rules preserve only the Hamiltonian. Take, from a
library, the four-qubit process $\ppar_i \tfrac18 \star Z_iZ_{i+1} \ppar
\ppar_i \tfrac1{16} \star X_i$, where $X_i$ applies $X$ to qubit $i$
and $Z_iZ_{i+1}$ applies $Z$ to qubits $i$ and $i+1$, composed with a
user's term $-\tfrac18 \star Z_2Z_3$ that cancels one of its terms. In every
Trotter step the two are separated by $X$-slices that anticommute with
them, so the circuit offers nothing to cancel. In the
program, rule~\eqref{rule:perm} and \cref{prop:inv} cancel them, and
the normal form needs \exCCnotNormal{} CNOT gates per step instead of
\exCCnotComposed, with less error at every level $n$:
\begin{center}
\begin{tabular}{r r r}
\toprule
$n$ & composed & normal form \\
\midrule
0 & 3.0975 & 2.1105 \\
2 & 0.8681 & 0.4997 \\
4 & 0.2186 & 0.1244 \\
\bottomrule
\end{tabular}

\end{center}

Rewriting can also make a circuit preserve a quantity exactly. Take
the six-qubit process whose adjacent pairs of qubits carry
$\tfrac1{32} \star (X_iX_{i+1} \ppar Y_iY_{i+1})$ and
$\tfrac1{64} \star Z_iZ_{i+1}$. The sum $X_iX_{i+1} + Y_iY_{i+1}$
exchanges $\ket{01}$ and $\ket{10}$ on its pair, so it preserves the
number of $1$s in a basis state, as do the $Z_iZ_{i+1}$ terms; but
$X_iX_{i+1}$ and $Y_iY_{i+1}$ alone do not. Scheduled by type, all $XX$
terms, then all $YY$, then all $ZZ$, the circuit started in
$\ket{010101}$ ends with a different number of $1$s with the
probability shown as the leak; scheduled pair by pair, a reordering by
rule~\eqref{rule:perm}, it preserves the count exactly at every level,
at nearly the same error:
\begin{center}
\begin{tabular}{r r r r r}
\toprule
$n$ & leak, by type & leak, by bond & error, by type & error, by bond \\
\midrule
0 & 1.7\% & 0 & 1.0232 & 1.0496 \\
2 & 0.1\% & 0 & 0.2512 & 0.2657 \\
4 & 0.0064\% & 0 & 0.0627 & 0.0665 \\
\bottomrule
\end{tabular}

\end{center}
This is the invariant reasoning of concurrent
programs~\cite{owickigries}: a transfer \texttt{a := a-1; b := b+1}
preserves $a + b$ in every interleaving only if it is atomic. Here a
block of consecutive commuting actions runs atomically in every step,
and the count is preserved when the block's sum commutes with
$M = \sum_i Z_i$; whether a schedule preserves it at every level is a
check on Pauli strings, made without any matrix. Merging the pair into
one action does the same by construction: by normal forms,
$\tfrac1{32} \star (X_iX_{i+1} \ppar Y_iY_{i+1})$ is equivalent to the
single action $u(\id \oplus \delta_4 \oplus \id \oplus
-\delta_4)u^\dagger$ with $u = \mathsf{CX} \cdot (\mathsf{H} \otimes
\mathsf{I})$, which turns both strings into diagonals at once.

Not every Hamiltonian requires approximation. A single diagonal interaction
commutes with itself, so its schedule is exact at every level, with no Trotter
error to refine away. The two-qubit interaction $Z \otimes Z$:
\begin{minted}{haskell}
zzParity :: Natural -> Proc
zzParity k = diag2 0 k k 0
\end{minted}
places a phase on the $\lvert01\rangle$ and $\lvert10\rangle$ states and
compiles to a single CNOT-phase-CNOT circuit whose unitary matches the ideal
exactly (Frobenius $= 0.0000$).
The contrast with \mintinline{haskell}{zxK} is the point. Refinement is the
machinery for the non-commuting case; where terms commute, the compiler produces
the exact evolution directly, and the schedule space collapses to a single
circuit.

These examples are deliberately minimal, chosen to isolate one behavior each. The
same pipeline, unchanged, compiles the multi-term molecular Hamiltonian of
\cref{s:applications}: a nine-term, two-qubit Pauli sum taken from a published
ion-trap experiment, whose Trotter schedule converges to the exact dynamics with
the same first-order rate demonstrated here.

\paragraph{Deciding Equivalence}
By full abstraction, $p \approx q$ is decided by comparing Pauli
representations, and the implementation computes them symbolically. A
bare action contributes the Walsh-Hadamard transform of its diagonal;
conjugation by a Clifford maps each Pauli string to a signed Pauli
string gate by gate, by the update rules of~\citet{aaronsongottesman};
coefficients stay exact dyadic rationals. Programs given as Pauli
strings keep the tensor shorthand of \cref{s:processes} symbolic and
have one string per action, so the check is linear in the program. The
table compares a family of Hamiltonians whose terms are the strings
$XZ\cdots ZX$ and $YZ\cdots ZY$ on up to four adjacent qubits, plus
single $Z$ terms, written as Pauli strings, with the Pauli gadgets a compiler emits for it (one $Z$, a ladder of
CNOT gates and a change of basis), on up to 256 qubits, where the
operator either program denotes is a $2^{256} \times 2^{256}$ matrix. A
gadget with a fault, the $\mathsf{S}$ of its $Y$ factors dropped, shows
up as two differing strings (column \emph{fault}):
\begin{center}
\begin{tabular}{r r r r r}
\toprule
qubits & terms & normal form & fault & time (ms) \\
\midrule
8 & 44 & 44 & 2 & 0.2 \\
16 & 100 & 100 & 2 & 0.7 \\
32 & 212 & 212 & 2 & 2.3 \\
64 & 436 & 436 & 2 & 7.9 \\
128 & 884 & 884 & 2 & 29.5 \\
256 & 1780 & 1780 & 2 & 126.7 \\
\bottomrule
\end{tabular}

\end{center}
Reorderings need no check, by rule~\eqref{rule:perm}; the check serves
every other rewrite.

\paragraph{Sanity Checking}

Leveraging a pre-existing implementation of $\Pi$ in Agda, we formalised
the core of \Hamilton, namely all the contents of Section~\ref{s:processes},
and the contents of Section~\ref{s:semantics} up to Proposition 14
over an abstract field and abstract matrices, as well as the first lemma of
Section~\ref{s:completeness}. The \emph{statements} of the other propositions in
Section~\ref{s:semantics} were also formalised. As usual, this caught a number of
minor infelicities in the ``paper math'' version.

The main obstacles were the many occurrences of ``$n$ times'' (and
variants thereof), whose degenerate case $n = 0$ is nonsensical, and the
definition of schedule equivalence, whose obvious transcription would put
defined functions into the indices of its type (what dependently typed
programmers call \emph{green slime}), making all subsequent reasoning
unbearably difficult.

We expect the main difficulty in formalizing the rest of
Sections~\ref{s:semantics} and~\ref{s:completeness} to lie in the
\emph{universality} theorems, which mix syntax and semantics, and in the
normalisation procedure, whose current proof uses ``without loss of
generality'' reasoning that will need to be made algorithmic.

\section{Application: Simulating a Hydrogen Bond}\label{s:applications}
\newcommand{\chemProgram}{p \;=\;\; & \tfrac{1024}{2^{12}} \star ZI \;\ppar\; \tfrac{752}{2^{12}} \star IZ \;\ppar\; \tfrac{-568}{2^{12}} \star IX \;\ppar\; \tfrac{448}{2^{12}} \star ZZ \;\ppar\; \tfrac{-305}{2^{12}} \star XX \\ \ppar\;\; & \tfrac{-305}{2^{12}} \star YY \;\ppar\; \tfrac{61}{2^{12}} \star XI \;\ppar\; \tfrac{-25}{2^{12}} \star ZX \;\ppar\; \tfrac{-5}{2^{12}} \star XZ}
\newcommand{\chemScale}{0.0192}
\newcommand{\chemUnitFs}{2.9}
\newcommand{\chemStepsNFour}{219}
\newcommand{\chemFreqs}{2323, 5260 and 11363}
\newcommand{\chemFloor}{2.8}
\newcommand{\chemDevNTwo}{0.17}
\newcommand{\chemDevNFour}{0.0042}
\newcommand{\chemCnotEmitted}{18}
\newcommand{\chemCnotCancelled}{10}
\newcommand{\chemCnotResynth}{3}

\citet{doi:10.1021/acs.jpclett.3c01601} used the QSCOUT trapped-ion
quantum computer to simulate the hydrogen bond of the molecule
DMANH\textsuperscript{+}. We summarise their experiment, then play the
part of their machine: we take their Hamiltonian as given, compile it
with \Hamilton, and compare the results with what they measured.

In DMANH\textsuperscript{+}, a proton sits between two nitrogen atoms;
each attracts it, and repels it at close range. A classical particle
under these forces would oscillate along the bond with a single
period; the quantum proton oscillates with several
periods. Discretizing the bond and using the mirror symmetry of the
molecule, the experiment tracks the proton at four positions $x_0,
\ldots, x_3$: at time $t$ it is in the superposition $a_0(t)\ket{x_0}
+ \cdots + a_3(t)\ket{x_3}$, whose amplitudes evolve under a
Hamiltonian $\tilde H$ that they publish. The experiment prepares the
proton at $x_3$, say, evolves it for times $t_0, \ldots, t_{N-1}$, and
measures its position, repeating each run to collect statistics. The
resulting series $|a_3(t_0)|^2, \ldots, |a_3(t_{N-1})|^2$ is a sum of
a few sinusoids whose frequencies are the gaps between the eigenvalues
of $\tilde H$; the three gaps from the lowest eigenvalue are
vibrational frequencies of the bond. A Fourier transform of the series
recovered them with a typical uncertainty of 3.3~cm$^{-1}$, the unit
chemists use for frequencies. To run the experiment, they needed a
\emph{small} quantum circuit. For each $t$, this circuit was computed on a
classical computer by calculating $e^{-i\tilde H t}$ and decomposing it
into an optimal circuit of three CNOT gates and seven single-qubit
rotations. This approach does not scale; for larger molecules, they
propose compiling the Hamiltonian term by term instead, by
Trotterisation, which is what \Hamilton does.

To write $\tilde H$ in \Hamilton, we turn each entry $h_{ij}$ of the
published $4 \times 4$ matrix into the term $h_{ij} \ket{x_i}\bra{x_j}$, so that
$\tilde H = \sum_{i,j} h_{ij} \ket{x_i}\bra{x_j}$. Two qubits store $x_i$
as the binary numeral of $i$, so that $x_3 = \ket{11}$, and the
identities
\[
  \ket{0}\bra{0} = \tfrac12(I + Z), \quad
  \ket{1}\bra{1} = \tfrac12(I - Z), \quad
  \ket{0}\bra{1} + \ket{1}\bra{0} = X, \quad
  \ket{0}\bra{1} - \ket{1}\bra{0} = iY
\]
turn each term into Pauli strings: a diagonal entry needs only $I$ and
$Z$, and an off-diagonal entry, together with its mirror image
$h_{ji} = h_{ij}$, has an $X$ or a $Y$ at each bit in which $i$ and $j$
differ. For instance,
\begin{align*}
  \ket{x_3}\bra{x_3} &= \tfrac14(II - ZI - IZ + ZZ), \\
  \ket{x_2}\bra{x_3} + \ket{x_3}\bra{x_2} &= \tfrac12(IX - ZX), \\
  \ket{x_1}\bra{x_2} + \ket{x_2}\bra{x_1} &= \tfrac12(XX + YY),
\end{align*}
where $ZI$ stands for $Z \otimes I$. Collecting terms, dropping the
identity (which shifts all energies equally), and scaling and rounding
the weights to $k = 12$ fractional bits gives the process
\[
\begin{aligned}
\chemProgram,
\end{aligned}
\]
built from $\delta_0$, $\mathsf{H}$ and $\mathsf{S}$ as in
\cref{s:processes}. The weights of $ZI$, $IZ$ and $ZZ$ come from the
diagonal of $\tilde H$, and the other six from its off-diagonal entries.
Each term $\frac{m}{2^k} \star P$ compiles to a single action.

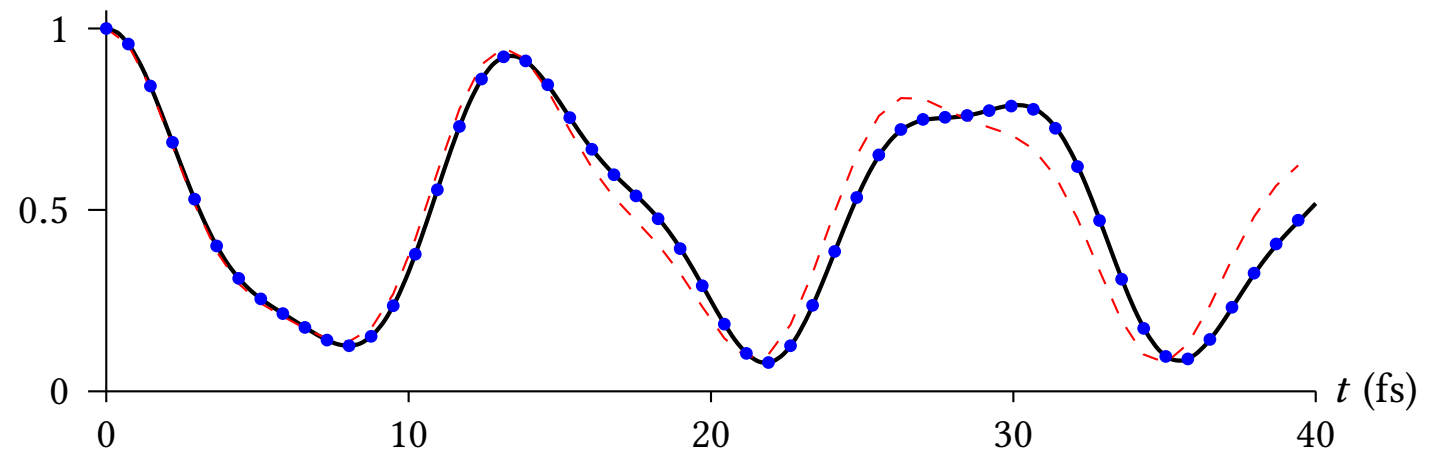
\begin{figure}
  \centering
  \begin{tikzpicture}[x=2mm, y=24mm, every node/.style={font=\footnotesize}]
    \draw (0,0) -- (40,0) node[right] {$t$ (fs)};
    \draw (0,0) -- (0,1.05);
    \foreach \tv in {0,10,20,30,40} \draw (\tv,0) -- (\tv,-0.03) node[below] {$\tv$};
    \foreach \pv in {0,0.5,1} \draw (0,\pv) -- (-0.6,\pv) node[left] {$\pv$};
    \draw[thick] plot[smooth] file {generated/chem_traj_exact.table};
    \draw[dashed, red] plot file {generated/chem_traj_n2.table};
    \draw[blue] plot[only marks, mark=*, mark size=0.35mm]
      file {generated/chem_traj_n4.table};
  \end{tikzpicture}
  \Description{The probability of finding the proton at position x3, ket 11,
    after starting it there, over 40 femtoseconds. The exact curve oscillates between about
    0.1 and 1; the points for n equal to 4 lie on it, and the dashed curve for
    n equal to 2 departs from it in places.}
  \caption{Probability $|a_3(t)|^2$ of finding the proton at
    $x_3 = \ket{11}$ at time $t$, starting there: exact evolution under $\tilde H$ (solid), and compiled circuits
    for $p$ with $k = 12$ at refinement levels $n = 2$ (dashed) and $n = 4$
    (dots).}
  \label{fig:chem}
\end{figure}

\paragraph{Running the Program}
In \Hamilton, the snapshot at time $t = j/2^n$ runs $\frac{j}{2^n}
\star p$, whose schedule is $j$ repetitions of one Trotter step: the
nine actions in sequence, each for $2^{-n}$ units of time. A
simulation is thus a loop around one short circuit, with a snapshot
after each iteration, and two numbers control its accuracy: the
precision $k$ of the weights, which fixes the rounding error, and the
refinement level $n$, which fixes the Trotter error. No unitary is
computed, and a step grows with the number of terms rather than with
the dimension. At two qubits the price is circuit size: a step has
\chemCnotEmitted{} CNOT gates as emitted and \chemCnotCancelled{} once
adjacent pairs cancel, so the \chemStepsNFour{} steps of
\cref{fig:chem} ($n = 4$) add up to more than two thousand, where the
experiment needs three for any $t$. Smaller circuits for the same
unitary are reachable by rewriting alone: the equational theory of
$\Pi_k$ is complete, so every $\Pi_k$ circuit with the same unitary is
derivable from ours, and cancelling CNOT pairs is the simplest such
derivation. The experiment's circuits are not among them: they
implement the exact evolution rather than our rounded, Trotterised
one, and their rotation angles, computed numerically, are in general
not dyadic.

We computed the outputs of the compiled circuits exactly, without
hardware noise. The eigenvalues of a compiled step determine the
frequencies at which repeated steps oscillate, which are the ones a
Fourier transform of the snapshots finds. The table gives the largest
error in the three frequencies:
\begin{center}
\begin{tabular}{r r r r}
\toprule
& \multicolumn{3}{c}{frequency error (cm$^{-1}$)} \\
\cmidrule(l){2-4}
$n$ & $k = 8$ & $k = 10$ & $k = 12$ \\
\midrule
1 & 243.0 & 225.5 & 220.2 \\
2 & 72.8 & 70.3 & 59.8 \\
3 & 23.3 & 25.4 & 13.3 \\
4 & 23.6 & 13.8 & 2.5 \\
8 & 25.3 & 9.9 & 2.8 \\
$\infty$ & 25.3 & 9.9 & 2.8 \\
\bottomrule
\end{tabular}

\end{center}
Down each column, refinement removes the Trotter error until only the
rounding error of the bottom row, the exact evolution of $p$, remains;
more bits lower it. \Cref{fig:chem} shows the same convergence along a
trajectory: $|a_3(t)|^2$ departs from the exact curve by up to
\chemDevNTwo{} at $n = 2$, and by at most \chemDevNFour{} at $n = 4$.
At $k = 12$ and $n \geq 4$ the frequency error is at most
\chemFloor~cm$^{-1}$, below the experiment's 3.3~cm$^{-1}$: compiling
term by term costs no accuracy that the experiment could detect. The two
numbers measure different things, though. Ours is the error of the
algorithm, computed exactly; theirs is the noise of a physical
measurement, on circuits with no algorithmic error. On hardware our
longer circuits would add noise, so at two qubits the experiment's
compiler is the better choice; ours is for molecules whose unitaries are
too large to compute.

\section{Related Work}\label{s:related}

\paragraph{Languages and compilers for Hamiltonians}
SimuQ~\cite{pengetal:simuq} is closest in purpose. A program in its
Hamiltonian Modeling Language describes the system to be simulated,
hardware providers describe what their machines can do in an abstract
analog instruction set, and a solver finds a pulse schedule that makes an
analog machine follow the program. SimuQ thus targets machines that are
programmed through their own Hamiltonian, where \Hamilton targets
circuits; it has no equational theory, and its notion of correctness is
that of its solver.
Paulihedral~\cite{li:paulihedral} compiles from an intermediate
representation of blocks of weighted Pauli strings, which is the normal
form of \Hamilton, through scheduling passes and passes that synthesize,
simplify and map the resulting circuits, for both near-term and
error-corrected machines. Its scheduling changes the order of terms
within a step, which rule~\eqref{rule:perm} certifies, and its outputs
can be checked against its inputs by normal forms
(\cref{s:implementation}).
Cobble~\cite{yuan:cobble} is a language for quantum linear algebra, with
simulation among its applications: programs manipulate matrices through
block encodings and are compiled to circuits, with resource analyses
and optimisations based on the quantum singular value
transformation~\cite{gilyenetal2019:qsvt}. A block encoding realises a
sum of matrices exactly at the cost of extra
qubits~\cite{lowchuang:qubitization}; \Hamilton approximates it with
none.
We claim no advantage in circuit size or fidelity over these systems;
what \Hamilton adds is an equational theory, complete for Hamiltonians,
under which their transformations can be certified.

\paragraph{Equational reasoning about Pauli rotations}
Phase gadgets, the ZX-calculus form of a rotation about a Pauli string,
support equational reasoning about circuits made of such rotations and
underlie gadget synthesis in the tket compiler~\cite{cowtanetal:gadgets}.
That reasoning concerns the unitaries of circuits at a fixed level.
\Hamilton reasons about the Hamiltonians they approximate, so its rules
also relate circuits whose unitaries differ, such as the reorderings and
cancellations of \cref{s:implementation}.

\paragraph{Calculi for quantum circuits}
$\Pi_k$~\cite{sqrtpi,freeq}, on which \Hamilton builds, is complete for
the unitaries of circuits; \Hamilton adds parallel composition and
halving on top of it and is complete for Hamiltonians
(\cref{s:completeness}).

\section{Conclusion}\label{s:conclusion}

A circuit is a sequence of gates; a Hamiltonian is a sum of interactions.
\Hamilton is built on the second reading, taking additive combination of
evolutions as primitive and recovering the sequential circuit only at compile
time, through the Lie-Trotter formula that is its operational semantics. The
result is a language in which Hamiltonians are first-class, with the metatheory
one expects of a circuit calculus: an equational theory that is fully abstract,
so process equivalence collapses to equality of dyadic-rational Pauli
coefficients; a universality theorem fixing the expressible Hamiltonians
exactly; and an execution model that is ordinary $\Pi_k$ compilation. Used as
a configurable simulator, its implementation compiles the published Hamiltonian
of a hydrogen bond into circuits that recover vibrational frequencies of the
bond to within the measurement uncertainty of the original experiment.
Simulation is one use among several: any computation programmed by a
Hamiltonian, from optimisation to eigenvalue estimation, can be written,
checked and compiled the same way.

The schedules \Hamilton compiles to are first-order Trotter, and
higher-order splittings~\cite{suzuki} and qubitised
simulation~\cite{gilyenetal2019:qsvt} the natural next step for
driving the residual error and fault-tolerant cost down. What most
separates the language from practice, though, is maturity: it has so
far been exercised only on small Hamiltonians, its circuits still await
optimisation (one Trotter step of the molecular example emits
\chemCnotEmitted{} CNOT gates, where two-qubit resynthesis needs
\chemCnotResynth), and connecting it to
the standard pipeline of electronic-structure packages and
fermion-to-qubit mappings, and scaling to the many-term operators of
real problems, is the work that adoption would require. One can
additionally imagine compiling not just to simulation algorithms, but
also to other algorithms where simulation is a key component, such as
quantum phase estimation.

A deeper direction is to widen the language itself. So far we have built
\Hamilton on the pure, reversible core of $\Pi_k$, but the full calculus is
richer: \citet{freeq} extend $\Pi_k$ with mid-computation measurement and with a
theory of irreversible, open-system quantum computation, and show that these
additions strengthen rather than disturb its completeness. Lifting the same
extensions to \Hamilton's additive layer would let it express the eigenvalues
and expectation values that estimation and variational algorithms return, and
reach the dissipative and finite-temperature dynamics that a purely unitary
language cannot.


\begin{acks}
RK was supported by the Independent Research Fund Denmark through
\emph{Sapere Aude: DFF-Research Leader} grant 5251-00024B, and the
Norwegian Research Council through grant 361350. We would like thank Javaria Ghafoor for finding and correcting a bug in the compiler.
\end{acks}

\subsection*{Data Availability Statement}

The Haskell / Python implementation and the Agda verification are
available at \url{https://doi.org/10.5281/zenodo.23675392}. The
accompanying artifact also contains commands that rebuild every table
and code listing the implementation.

\subsection*{Disclosure of Use in AI in Research}

The code accompanying this paper was written by hand, with Claude generating parts of it in response to requests for specific rewrites. No generative AI was used for theory or proofs. The authors take full responsibility for all content, including the code.

\bibliography{bibliography.bib}


\end{document}